\documentclass[12pt,twoside]{article} 

\usepackage{bu_ece_report}
\usepackage{graphicx}
\usepackage{balance}  
\usepackage{times}
\usepackage{epsfig}
\usepackage{amsmath}
\usepackage{amssymb}
\usepackage{url}
\usepackage{multirow}
\usepackage{authblk}
\usepackage{color}
\usepackage{ragged2e}
\usepackage{makecell}
\begin{document}

\buecedefinitions%
        {Understanding Big Data Analytic Workloads on Modern Processors}
        {}
        {Zhen Jia, Lei Wang, Jianfeng Zhan, Lixin Zhang, Chunjie Luo, Ninghui Sun}
        {Beijing, China}
        {April.\ 10, 2015}
        {2015-4} 

\buecereporttitlepage



\buecereportsummary{Big data analytics applications
play a significant role in data centers, and hence
it has become increasingly important 
to understand their
behaviors in order to further improve the performance of data center computer systems, in which characterizing representative workloads is a key practical problem.
In this paper, after investigating three most important application domains in terms of page views and daily visitors,
we chose 11 representative data analytics workloads and characterized their micro-architectural behaviors  by using hardware performance counters,
so as to understand the impacts and implications of data analytics
workloads on the systems equipped with modern superscalar out-of-order processors.
Our study reveals that big data analytics applications themselves
share many inherent characteristics,
which place them in a different class from traditional workloads and scale-out services.
To further understand the characteristics of big data analytics workloads we performed a correlation analysis of CPI (cycles per instruction) with other micro-architecture level characteristics
and an investigation of the big data software stack impacts on application behaviors.
Our correlation analysis showed that even though 
big data analytics workloads own notable pipeline front end stalls, the main factors affecting the CPI performance are long latency data accesses rather than the front end stalls.
Our software stack investigation found that the typical big data software stack significantly contributes to the front end stalls and incurs bigger working set.
Finally we gave
several recommendations for architects, programmers and big data system designers with the knowledge acquired from this paper.
}




\section{Introduction} \label{intro}

the context of digitalized information explosion,
more and more
businesses are analyzing massive amount of data -- so-called big data --
with the goal of converting big data to ``big value'' by means of modern data center systems.
Typically, data center workloads can be classified into two categories: services and data analytics workloads as mentioned in \cite{zhan2012high} and \cite{barroso2009datacenter}.
For the data analytics workloads always process a large mount of data in data centers, we call them big data analytics workloads. 
Typical big data analytics workloads include business intelligence, machine learning, bio-informatics, and ad hoc analysis~\cite{InfrastructureAtFacebook,HadoopUsageReport}.

The business potential of the big
data analytics applications is a driving force behind the design of
innovative data center systems including both hardware and software~\cite{zhancost,wang2012cloud,sang2012precise,jiacharacterization}.
For example, the recommendation system is a typical example with
huge financial implications, aiming at recommending
the right products to the right buyers by mining user behaviors and other logs.
Given that big data analytics is a very important application area, there is a need
to identify the representative data analytics algorithms or applications in big data fields
and understand their performance characteristics with the purpose of improving the big data analytics systems' performance~\cite{jia2014implications}.
In order to achieve this purpose, the following two questions should be answered.
1). What are the potential bottlenecks and optimization points with higher priority in current systems.
2). What programmers should pay attention to when they develop applications with modern software stacks in order to gain more efficient big data analytics applications.
This paper seeks to address the above questions by characterizing representative big data analytics applications.
\subsection{Big Data analytics Workloads}
In order to identify representative big data analytics applications in data centers, we single out three important application domains in Internet services:  \emph{search engine, social networks, and electronic commerce} (listed in Figure~\ref{share})
according to widely acceptable metrics --- the number of page views and daily visitors. And then, we choose eleven representative big
data analytics workloads (especially intersection workloads) among the three application domains.
Considering our community may feel interest in using
those workloads to evaluate the benefits of new system designs and implementations, we release those workloads and the corresponding data sets into an open-source big data benchmark suite---BigDataBench~\cite{wang2014bigdatabench,gao2013bigdatabench}, which is publicly available from~\cite{BigDataBench_homepage}.
Based on selected representative big data analytics applications, we embark on a study to understand big data analytics workloads' behaviors on modern processors.
We first characterize big data analytics workloads with a more pragmatic experiment approach
in comparison with that of CloudSuite described in~\cite{ferdman2011clearing}.
We adopt larger input data sets varying from 147 to 187 GB that are stored in both the memory and disk systems instead of completely storing data (only 4.5GB for \emph{Naive Bayes} in~\cite{ferdman2011clearing}) in the memory system. And for each workload, we collect the performance data of
the whole run time after the warm-up instead of a short period (180 seconds in~\cite{ferdman2011clearing}).

We find that big data analytics applications share many inherent characteristics, which place them in a different class from desktop (SPEC CPU2006), HPC (HPCC), traditional service  (SPECweb2005 and TPC-W), chip multiprocessors (PARSEC), and scale-out service (four among six benchmarks in ClousSuite paper~\cite{ferdman2011clearing}) workloads.
Meanwhile the service workloads in data center (scale-out service workloads) share many similarities in terms of micro-architecture characteristics with that of traditional service workloads, so in the rest of this paper, we just use the service workloads to describe them.
Furthermore, we perform a correlation analysis between cycles per
instruction (CPI) performance and micro-architecture characteristics to find the potential optimization methods for big data analytics workloads.
At last we analyze the impacts of a typical big data software stack on critical metrics that have proved by correlation analysis for big data analytics applications on modern processors as a case study and show the aspects that programmers should pay much attention to.

\subsection{Paper Contributions and Outlines}
This paper has the following major contributions:

1) The characterization of big data analytics workloads and comparison with traditional workloads.
Base on the characterization we find that:
\begin{itemize}
\item The big data analytics workloads have higher instruction level parallelism (i.e. IPC) than that of the services workloads while lower than those of computation-intensive HPCC workloads, e.g., \emph{HPC-HPL, HPC-DGEMM} and  chip multiprocessors workloads.
\item Corroborating previous work \cite{ferdman2011clearing}, both the big data analytics workloads and  service workloads suffer from notable pipeline front end stalls.
\item The significant differences between the big data analytics workloads and the service workloads (four among six benchmarks
in ClousSuite~\cite{ferdman2011clearing}, SPECweb and TPC-W) in terms of processor pipeline stall breakdown: the big data analytics workloads suffer more stalls in the out-of-order part of the pipeline (about 57\% on average), while the service workloads
suffer more stalls before instructions entering the out-of-order part of pipeline (about 75\% on average).
\item Big data analytics workloads 
have lower L2 cache miss ratios (about 11 L2 cache misses per thousand instructions
on average) than those of the service workloads (about 66 L2 cache misses per thousand instructions on average) while higher than those of the HPCC workloads.
Meanwhile, for the big data analytics and service workloads, on the average 85.5\% and 95.5\% of L2 cache misses are hit in L3 cache (last level cache), respectively.
For the service workloads, our observations corroborate the previous work~\cite{ferdman2011clearing}: the L2 cache is ineffective.
\item For the big data analytics workloads, the misprediction
ratios are lower than those of most of the service workloads, which implies that the branch predictor
of modern processor is good. Further more, a simpler branch predictor may be preferred so as to save power and die area for big data analytics workloads.
\end{itemize}

2) The correlation analysis of the CPI performance and other micro-architecture metrics. Our results reveal that:
\begin{itemize}
\item
Although big data analytics workloads own a notable processor pipeline front end stall in our workload characterization study (Section~\ref{characterization_BDbenchmarks}), the front end stall does not have a strong correlation with CPI performance.
This implies that the front end stall is not the factor that affects CPI performance most for data analytics workloads from the perspective of micro-architecture.
\item
For big data analytics workloads, the TLB and private united cache (L2 cache in our architecture) performances have strong correlations with CPI performance. So the TLB and private united cache
need to be optimized with the highest priority in order to achieve better performance.
Further more, considering our findings in workload characterization study (Section~\ref{characterization_BDbenchmarks}), the L2 cache miss ratio is acceptable for big data analytics workloads and
the last level cache can satisfy most of cache misses from previous level caches.
So reducing the capacity of last level cache properly may benefit the performance,
since a smaller last level cache can shorten last level cache hit latency and reduce L2 cache miss penalty, which corroborates previous work~\cite{ferdman2011clearing,lotfi2012scale}.
Moreover, for modern processors dedicate approximately half of the die
area to caches, a smaller last level cache can also improve the energy-efficiency of processor and save the die size.
\end{itemize}
3) The investigation of modern big data software stack's impacts on big data analytics application behaviors from the perspective of micro-architecture. We find that the big data software stack has impacts on the following aspects:
\begin{itemize}
\item The big data software stack makes contribution to front end stalls
by increasing application's binary size and further increases the pressure on instruction fetch unit.

\item The big data software stack incurs larger working set and prolongs the memory access latency, especially for load operations.

\item 
Most of the big data software stack functions are
implemented on user-mode and do not invoke many system calls. The large amounts of user-mode instructions reduce the kernel-mode instruction ratio of the whole application.

\end{itemize}

The remainder of the paper is organized as follows.
Section~\ref{related} lists the related work.
Section~\ref{character} states our experiment methodology.
Section~\ref {characterization_BDbenchmarks} presents the micro-architectural characteristics of the data analysis workloads in comparison with other benchmark suites.
Section~\ref{corr} analyzes the correlation of each of the measured characteristics with CPI.
Section~\ref{software} investigates a typical big data analytics software stack's impacts on application behaviors.
Section \ref{conclusion} draws conclusions of the full paper. 

\section{Related Work} \label{related}
There have been much work proposed to evaluate data mining algorithms
or evaluate clusters using data analytics workloads in different aspects,
such as~\cite{narayanan2006minebench,huang2010hibench,awasthi2015system,anwar2014use,ferdman2011clearing} and etc.
Narayanan et al.~\cite{narayanan2006minebench} characterize traditional data analytics workloads on single node other then workloads running at data center scale.
Huang et al.~\cite{huang2010hibench} characterize the MapReduce framework in system
level performance. They evaluate the Hadoop framework and
do not focus on the micro-architecture's characteristics.
Awasthi et al.~\cite{awasthi2015system} also perform a system level characterization of data center applications.
Anwar et al.~\cite{anwar2014use} conduct a quantitative study of representative Hadoop applications on five hardware configurations with the purpose of evaluating the different clusters' performance.
The state-of-the-art work of characterizing scale-out (data center) workloads on a micro-architecture level is CloudSuite~\cite{ferdman2011clearing}.
However, CloudSuite paper is biased towards online service workloads: among six benchmarks, there are four scale-out service workloads, (including \emph{Data Serving, Media Streaming, Web Search,  Web Serving}), and only one big data analytics workload---{\em Naive Bayes}. One application can not cover all the characteristics big data analytics applications own.
Our work shows that the data analytics workloads are significantly diverse in terms of
micro-architectural level characteristics (Section~\ref{characterization_BDbenchmarks}) on modern processors.
Previous work also finds that big data analytics applications
show varying performance, energy behavior and preferable system configuration parameters~\cite{li2014mronline,krish2014varphisched,iqbal2014towards}. %
In a word, only one application is not enough to represent various categories of big data analytics workloads.

\begin{table}[hbtp]
\caption{Representative big data analytics workloads}\label{workloads}
\centering  \scriptsize \sffamily
\begin{tabular}{|c|c|c|c|c|} \hline
No. & Workload & Input Data Size    & \#Retired Instructions (Billions)  & Source \\ \hline
 1 & Sort      & 150 GB  documents & 4578 &Hadoop example \\ \hline
 2 & WordCount  & 154 GB documents  & 3533 &Hadoop example\\ \hline
 3 & Grep   & 154 GB  documents & 1499 & Hadoop example \\ \hline
 4 & Naive Bayes  & 147 GB text  & 68131& Mahout\cite{mahout} \\ \hline
 5 & SVM  & 148 GB html file    & 2051 &our implementation\\   \hline
 6 & K-means  & 150 GB vector  & 3227 &Mahout\\ \hline
 7 & Fuzzy K-means & 150 GB vector &15470 & Mahout  \\ \hline
 8 & IBCF  & 147 GB ratings data & 32340 & Mahout \\  \hline
  9& HMM   & 147 GB html file  &1841  &our implementation \\  \hline
 10 & PageRank  & 187 GB web page & 18470  &Mahout\\  \hline
 11 & Hive-bench & 156 GB  DBtable  & 3659 &Hivebench\\ \hline
\end{tabular}
\end{table}

\begin{table}[hbtp]
\caption{Scenarios of big data analytics applications.}\label{algorithm_scenarios}
\centering \scriptsize \sffamily
\begin{tabular}{|c|c|c|} \hline
  Name& Domain &Scenarios \\ \hline
          &search engine     &  	Log analysis \\ 
 	Grep    & social network  &Web information extraction \\ 
 	        & electronic commerce  &Fuzzy search  \\ \hline
  Bayes & social network  & Spam recognition\\ 
             & electronic commerce	& Web page classification \\ \hline
           & social network      & Image Processing \\ 
 	SVM       & electronic commerce   & Data Mining \\ 
 	          &      & Text Categorization \\ \hline
  PageRank &search engine & Compute the page rank \\ \hline
  Fuzzy  &search engine    & Image processing \\ 
  K-means, & social network &High-resolution landform\\
  K-means        & electronic commerce & classification \\ \hline 
         & social network   &  	Speech recognition \\ 
   HMM   & search engine  & 	Word Segmentation \\ 
          &  & 	Handwriting recognition \\ \hline
             &search engine   & Word frequency count \\ 
  WordCount & social network   & Calculating the TF-IDF value \\ 
 	             & electronic commerce &Obtaining the user operations count\\ \hline
  Sort & electronic commerce & Document sorting\\ 
	     & search engine & Pages sorting \\
       & social network &  \\ \hline
\end{tabular}
\end{table}

\section{Experimental Setup}
\label{character}

This section firstly describes the experimental environments on which we conduct our study,
and then explains our experiment methodology.





\subsection{Workloads Selection}  \label{workloads_choice}
\begin{figure}
\centering
\includegraphics[scale=0.8]{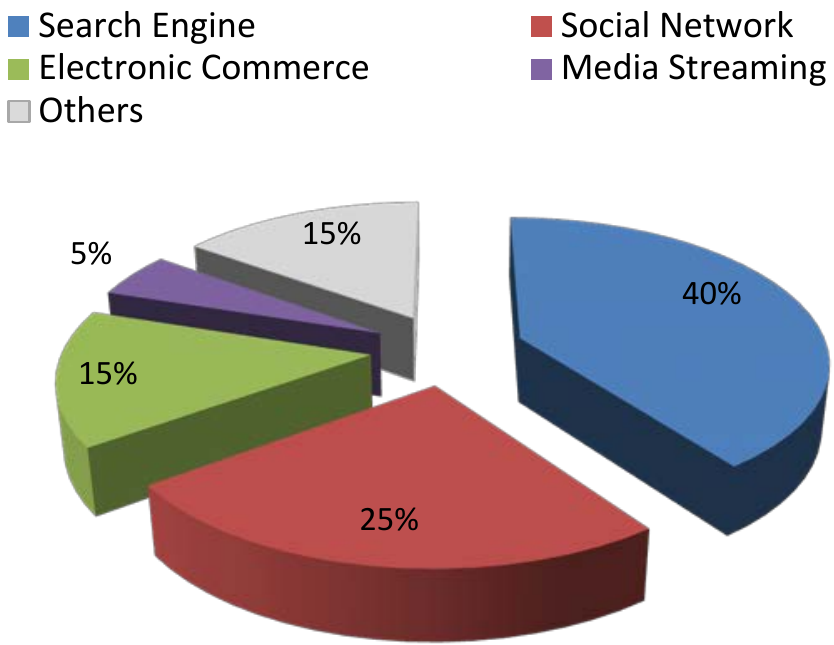}
\caption{Top sites in the web \cite{Alexa}.}\label{share}
\end{figure}

In order to find representative big data analytics workloads 
we firstly decide and rank the main application domains according to widely
acceptable metrics---the number of pageviews and daily visitors,
and then single out the main applications from the most important application domains.
We investigate the top sites listed in Alexa~\cite{Alexa},
of which the rank of sites is calculated using a combination of average daily visitors and page views. 
We classified the top 20 sites into 5 categories including search
engine, social network, electronic commerce, media streaming and others.
Figure~\ref{share} shows the categories and their respective share. To keep concise and simple, we focus on
the top three application domains: \emph{search engine}, \emph{social networks} and \emph{electronic commerce}.

We choose the most popular applications in those three application domains. Table~\ref{algorithm_scenarios} shows application scenarios of each workload, which is characterized in this paper, indicating most of our chosen workloads are intersections among three domains.
Considering our community may feel interest in using those
those workloads to evaluate the benefits of new system designs and implementations, we release those workloads and corresponding data sets into an open-source big data benchmark suite---BigDataBench~\cite{wang2014bigdatabench,ming2014bdgs,zhu2014bigop},
which is an open-source big data benchmark suite modeling diversified typical and important big data application domains.

\subsection{Hardware Configurations} \label{hwcon}

We use a 5-node Hadoop cluster (one master and four slaves) to run all big data analytics workloads.
The nodes in our Hadoop cluster are connected through 1 Gb ethernet network.  Each
node has two Intel Xeon E5645 (Westmere) processors and 32 GB memory.
A Xeon E5645 processor includes six physical  out-of-order cores with
speculative pipelines. Each core has private L1 and L2
caches, and all cores share the L3 cache. Table~\ref{hwconfigeration}
lists the important hardware configurations of the processor.

\begin{table}
\caption{Details of Hardware Configurations.}\label{hwconfigeration}
\center \scriptsize \sffamily
\begin{tabular}{|c|c|}
  \hline
  CPU Type & Intel \textregistered Xeon E5645\\ \hline
  \# Cores & 6 cores@2.4G \\ \hline
  \# threads& 12 threads \\ \hline
	\#Sockets & 2 \\ \hline
  \hline
  ITLB & 4-way set associative, 64 entries \\ \hline
  DTLB & 4-way set associative, 64 entries \\ \hline
  L2 TLB& 4-way associative, 512 entries \\ \hline
  L1 DCache & 32KB, 8-way associative, 64 byte/line \\ \hline
  L1 ICache & 32KB, 4-way associative, 64 byte/line \\ \hline
  L2 Cache & 256 KB, 8-way associative, 64 byte/line \\ \hline
  L3 Cache &  12 MB, 16-way associative, 64 byte/line \\ \hline
  Memory & 32 GB , DDR3 \\  \hline
\end{tabular}
\end{table}

\subsection{Big Data Analytics Applications Setups} \label{configuration}
All the big data analytics applications are implemented on the  Hadoop~\cite{white2009hadoop} system,
 which is an open source MapReduce implementation. 
The version of Hadoop and JDK is 1.0.2 and 1.6.0, respectively.
For data warehouse workloads, we use Hive of the 0.6 version.
Each node runs Linux CentOS 5.5  with the 2.6.34 Linux kernel.
Each slave node is configured with 24 map task slots
and 12 reduce task slots. 
For each map and reduce task, we assigned 1 GB Java heap in order to achieve better performance.

Table~\ref{workloads} presents the size of input data set and the instructions retired of each big data analytics workload. The input data size varies from 147 to 187 GB. In comparison with  that of CloudSuite described in~\cite{ferdman2011clearing}, our approach are more pragmatic. We adopt a larger data input that are stored in both memory and disk systems instead of completely  storing data (only  4.5 GB for \emph{Naive Bayes} in~\cite{ferdman2011clearing}) in memory. The number of instructions retired of the big data analytics workloads
ranges from thousand of billions to tens of thousands of billions, which indicates that those applications
are not trivial ones.

\subsection{Compared Benchmarks Setups} \label{traditionalben}
In addition to big data analytics workloads, we
deployed several
 benchmark suites, including SPEC CPU2006, HPCC, PARSEC, TPC-W, SPECweb 2005,  and CloudSuite---a scale-out benchmark suite for cloud computing~\cite{ferdman2011clearing}, and compared them with big data analytics workloads.

\subsubsection{Traditional benchmarks setups}
SPEC CPU2006: we run the official applications with the first reference input, reporting results averaged into two groups, integer benchmarks (\emph{SPECINT}) and floating point benchmarks (\emph{SPECFP}). The gcc which we used to compile the SPEC CPU is version 4.1.2.

	HPCC: we deploy HPCC --a representative HPC benchmark suite. The HPCC version is 1.4. It has seven benchmarks\footnote{\emph{HPL} solves linear  equations. \emph{STREAM} is a simple synthetic benchmark, streaming access memory. \emph{RandomAccess}  updates (remote) memory randomly. \emph{DGEMM} performs matrix multiplications.
\emph{FFT} performs discrete fourier transform. \emph{COMM} is a set of tests to measure latency and bandwidth of the interconnection system.}, including \emph{HPL}, \emph{STREAM}, \emph{PTRANS}, \emph{RandomAccess}, \emph{DGEMM}, \emph{FFT}, and \emph{COMM}.
We run each benchmark respectively.
	
SPECweb 2005: we run the bank application as the Web server on one node with 24 GB data set.
We use distributed clients to generate the workloads, and the number of the total simultaneous sessions is 3000.

PARSEC: we deploy PARSEC 2.0 Release. We run all benchmarks with native input data sets and use gcc with version 4.1.2 to compile them.

TPC-W: we deploy a Java TPC-W Implementation Distribution from University of Wisconsin-Madison~\cite{tpc-w} with MySQL version  5.1.73  and JDK version 1.6.0.


\subsubsection{CloudSuite Setups}

CloudSuite 1.0 has six benchmarks, including one big data analytics workload---
\emph{Naive Bayes}. We also choose \emph{Naive Bayes} as one of the
representative big data analytics workloads with a larger data input set (147 GB). In~\cite{ferdman2011clearing}, the data input size is only 4.5 GB.

We set up the other five  benchmarks following the introduction on the CloudSuite web site~\cite{cloudsuite}.
		
		Data Serving: we benchmark \emph{Cassandra} 0.7.3 database with 30 million records. The request is generated by a YCSB~\cite{cooper2010benchmarking} client with a 50:50 ratio of read to update.
		
		Media Streaming: we use \emph{Darwin} streaming server 6.0.6. We set 20 Java processes and issue 20 client threads by using the Faban driver~\cite{faban} with GetMediumLow 70 and GetshortHi 30.
		
		Software Testing: we use the \emph{cloud9} execution engine, and run the printf.bc coreutils binary file.
		
		Web Search: we benchmark a distributed  \emph{Nutch} 1.1 index server. The index and data segment size is 17, and 35 GB, respectively.
		
		Web Serving: we characterize a front end of \emph{Olio} server. We simulate 500 concurrent users to send requests with 30 seconds ramp-up time and  300 seconds steady state time.

\subsection{Experimental Methodology} \label{eMethodology}
Modern superscalar Out-of-Order (OoO) processors prevent us from breaking down the execution time precisely
due to overlapped work in the pipeline~\cite{ferdman2011clearing,keeton1998performance,eyerman2006performance}.
The retirement centric analysis is also difficult to
account how the CPU cycles are used because the pipelines will continue executing instructions even though the
instruction retirement is blocked~\cite{levinthal18027cycle}.
So in this paper, we focus on counting  
\emph{cycles stalled due to resource conflict}, e.g. the
reorder buffer full stall, which prevents new instructions from entering the pipelines.

We get the micro-architectural data by using hardware performance counters to measure the architectural events. In order to monitor micro-architectural events, a Xeon processor provides several performance event
select MSRs (Model Specific Registers), which specify hardware events to be counted, and performance monitoring counter MSRs, which store results of performance monitoring events.
We use Perf---a profiling tool for Linux 2.6+ based systems~\cite{perf}, to manipulate those MSRs by specifying the event numbers and corresponding unit masks.
We collect about 20 events whose number and corresponding unit masks can be found in the Intel Software Developer's Manual~\cite{intelref}.
In addition, we access the \emph{proc} file system to collect OS-level
performance data, such as the number of disk writes. 


We perform a ramp-up period for each application, and then start collecting the
performance data.
Different from the experiment methodology of CloudSuite, which only performs 180-second measurement, the performance data we collected cover the whole lifetime of each application, including map, shuffle, and reduce stages.
We collect the data of all the four working nodes and report the
mean value.



\section{Characterization Results}
\label{characterization_BDbenchmarks}


We provide a detailed analysis of
the inefficiencies of running big data analytics workloads on modern OoO (Out of Order) processors in the rest of this section.

\subsection{Instructions Execution} \label{sub_ipc}

\begin{figure}
\centering
\includegraphics[scale=0.75]{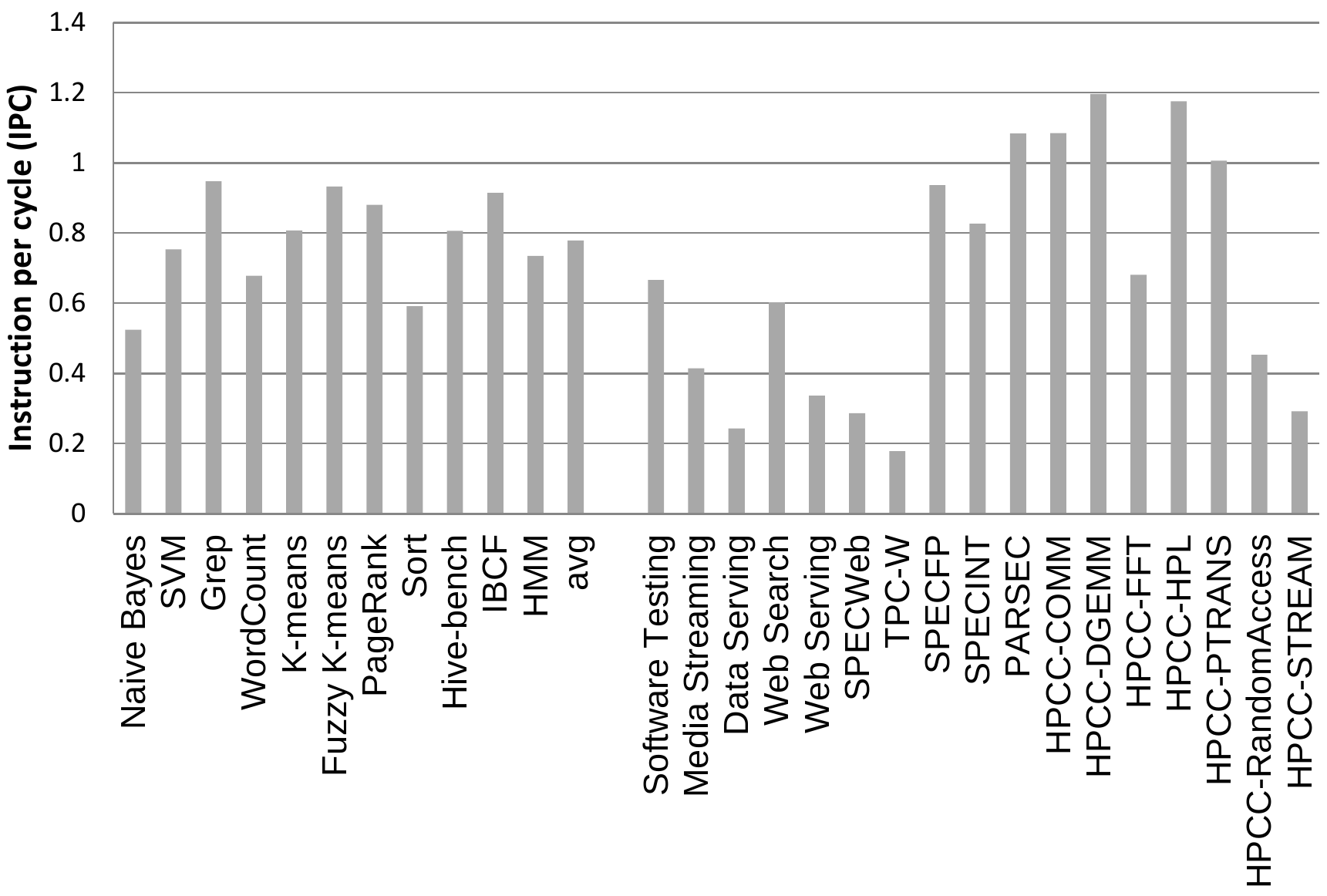}
\caption{Instructions per cycle for each workload.}\label{ipc}
\end{figure}


Instructions per cycle (in short IPC) is used to measure instruction level parallelism,
indicating how many instructions can execute simultaneously.
Our processors have 6 cores,  and each core can commit up to 4 instructions on each
cycle in theory.
However, for different workloads, IPC can be limited by pipeline stalls and
data or  instructions dependencies.

Figure \ref{ipc} shows IPC  of each workload.
The CloudSuite has six benchmarks, among which we report the Naive Bayes on the leftmost side, separated from the other five workloads (in the middle side), since \emph{Naive Bayes} is also included into our eleven workloads.

The main workloads of CloudSuite (four among six) are service workloads: \emph{Media Streaming, Data Services, Web Services, and Web Search}. From Figure \ref{ipc}, we can observe that the service workloads, including four of CloudSuite, \emph{TPC-W} and \emph{SPECweb}, have the low IPCs (all less than 0.6) in comparison with
the other workloads, including our chosen big data analytics workloads, \emph{PARSEC}, \emph{SPECFP}, \emph{SPECINT}, and most of HPCC workloads.

Most of big data analytics workloads have middle IPC values, greater than those of the service workloads.
The IPCs of the eleven big data analytics workloads ranges from 0.52 to 0.95 with an average value of 0.78. The avg bar in Figure \ref{ipc} means the average IPC of the eleven big data analysis workloads. \emph{Naive Bayes} has the lowest IPC value among the eleven big data analysis workloads.
The IPCs of the HPCC workloads have a large discrepancy
among each workload since they are all micro-benchmark designed for measuring different aspects of systems. For example, \emph{HPCC-HPL} and \emph{HPCC-DGEMM} are computation-intensive, and hence have a higher IPC (close to 1.2). While  \emph{HPCC-STREAM} is designed to stream access memory,  it has poor temporal locality, causing long-latency memory accesses, and hence it has lower IPCs (less than 0.5).


\begin{figure}
\centering
\includegraphics[scale=0.7]{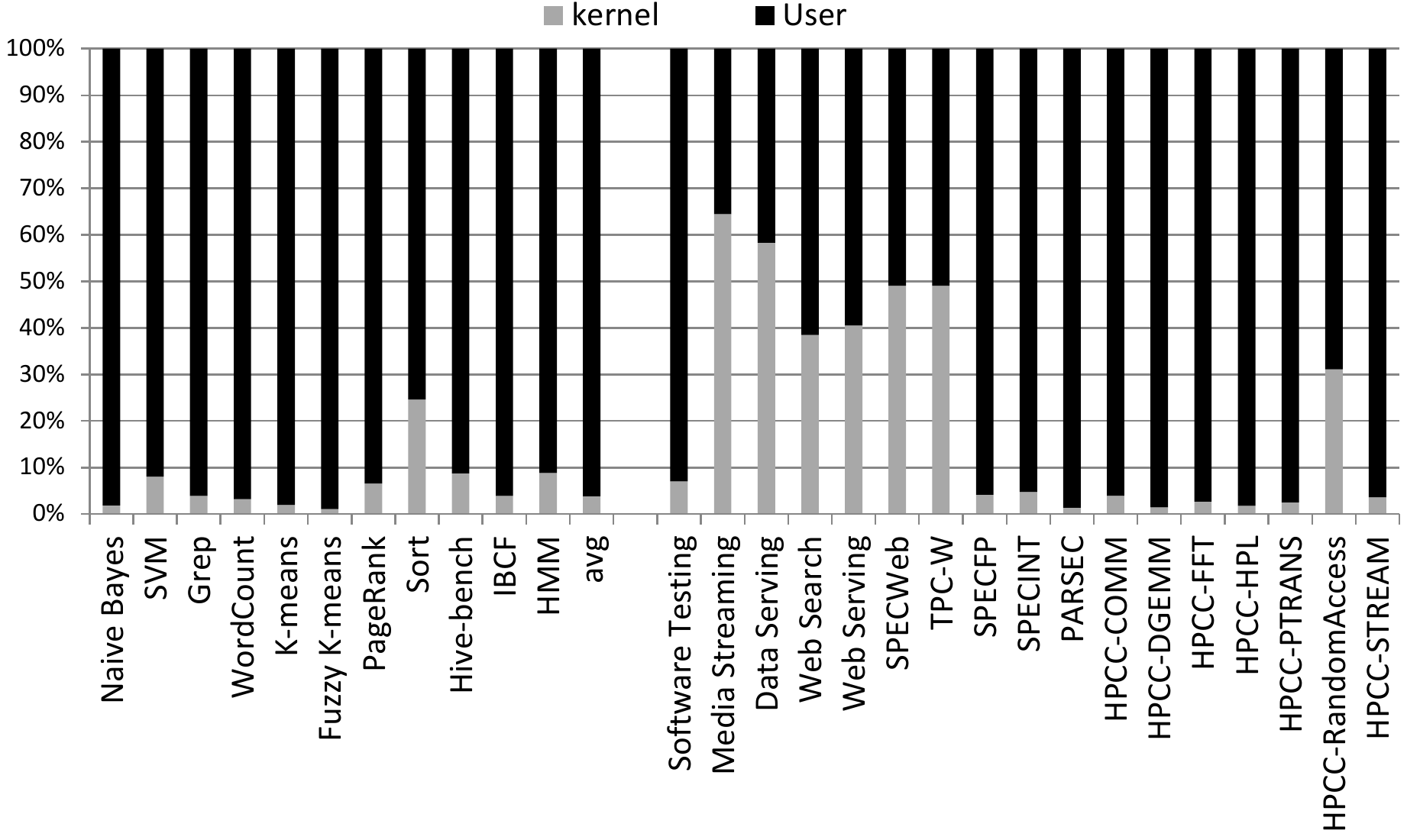}
\caption{User and Kernel Instructions Breakdown.}\label{ins_share}
\end{figure}

Figure \ref{ins_share} illustrates the retired instructions breakdown of each workload.
We also notice that the service workloads (four of CloudSuite, \emph{TPC-W }\emph{SPECWeb}) execute a large percentage of kernel-mode instructions (greater than 40\%), while most big data analytics workloads execute a small percentage of kernel-mode instructions.
The service workloads have higher percentages of kernel-mode instructions because serving a large amount of requests will result in a large number of network and disk activities.



\begin{figure}
\centering
\includegraphics[scale=0.75]{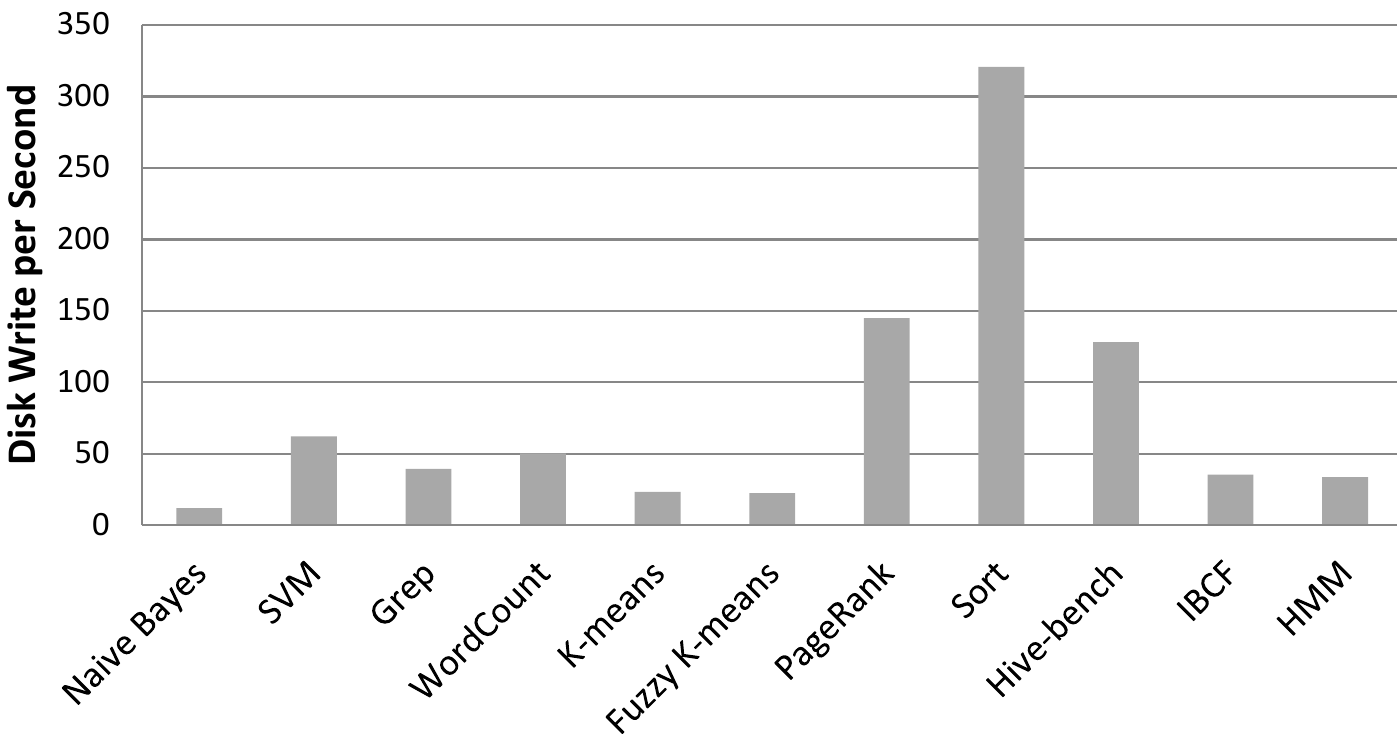}
\caption{Disk Writes per Second.}\label{diskwrite}

\end{figure}

Among the big data analytics workloads, only \emph{Sort} has a high proportion (about 24\%) of kernel-mode instructions whereas on average the big data analytics workloads only have about 4\% instructions executed in kernel-mode.
This is caused by the two unique characteristics of \emph{Sort}.
The first one is that different from most of the big data analytics workloads,
the input data size of \emph{Sort} is equal to the output data size.
So in each stage of the MapReduce job, the system will write a large amount of output data to local disks or transfer a large amount of data over network.  This characteristic makes \emph{Sort} have more I/O operations than other workloads.
The second unique characteristic is that \emph{Sort} has simple computing logic, only comparing. So it can process a large amount of data in a short period of time.
Those characteristics let \emph{Sort}  involve more frequent I/O operations (both disk and network).
So in comparison with other big data analytics workloads, \emph{Sort} is much OS-intensive.
Figure~\ref{diskwrite} depicts disk writes per second of each big data analytics workload.
We can find that \emph{Sort} has the highest disk writes frequency.
We also observed that network communication activities of \emph{Sort} are also more frequent than those of the other big data analytics workloads.

Among the HPCC workloads, \emph{RandomAccess} has a large percentage of kernel-mode instructions (about 31\%). \emph{RandomAccess} measures the rate of integer random updates of (remote) memory.
 An update is a read-modify-write operation on a table of 64-bit words, and it involves a large amount of  $copy\_user\_generic\_string$ system calls.
 The other factors contributing a large percentage of kernel-mode instructions need further investigations.

\textbf{\emph{Observations}}:

Big data analytics workloads have higher IPCs than those of services workloads, which are characterized by CloudSuite, traditional web server workload (\emph{SPECweb2005}) and traditional transactional web workload (\emph{TPC-W}), while lower than those of computation-intensive workloads, e.g., \emph{HPCC-HPL, HPCC-DGEMM, PARSEC}.
Meanwhile we also observe that the most of big data analytics workloads involve less kernel-mode instructions than that of the service workloads.

\subsection{Pipeline Behaviors}\label{rrs}
Modern processor implements dynamic execution using out of order and speculative engine.
The whole processor architecture can be divided into two parts: including an in-order front
end, which fetches, decodes, and issues instructions, and an out-of-order back end, which
executes instructions and write data back to register files.
A stall can happen in any part of the pipeline.
In this paper we focus on the major pipeline stalls (\emph{not exhausted}),   including front end (instruction fetch), register
allocation table (in short \emph{RAT}), load-store buffers, reservation station (in short \emph{RS}), and re-order buffer (in short \emph{ROB}).
For modern X86 architecture, front end will fetch instructions from L1 Instruction cache and then decode the CISC
instructions into RISC-like instructions, which  Intel calls micro-operations.
\emph{RAT} will change the registers used by the program
into internal registers available. 
Load-store buffers are also known as memory order buffers, holding in-flight memory
micro-operations (load and store), and they ensure that writes to memory take place in the
right order.
\emph{RS} queues micro-operations until all source operands are ready.
\emph{ROB} tracks all micro-operations in-flight and make the out-of-order executed
instructions retire in order.

\begin{figure}
\centering
\includegraphics[scale=0.75]{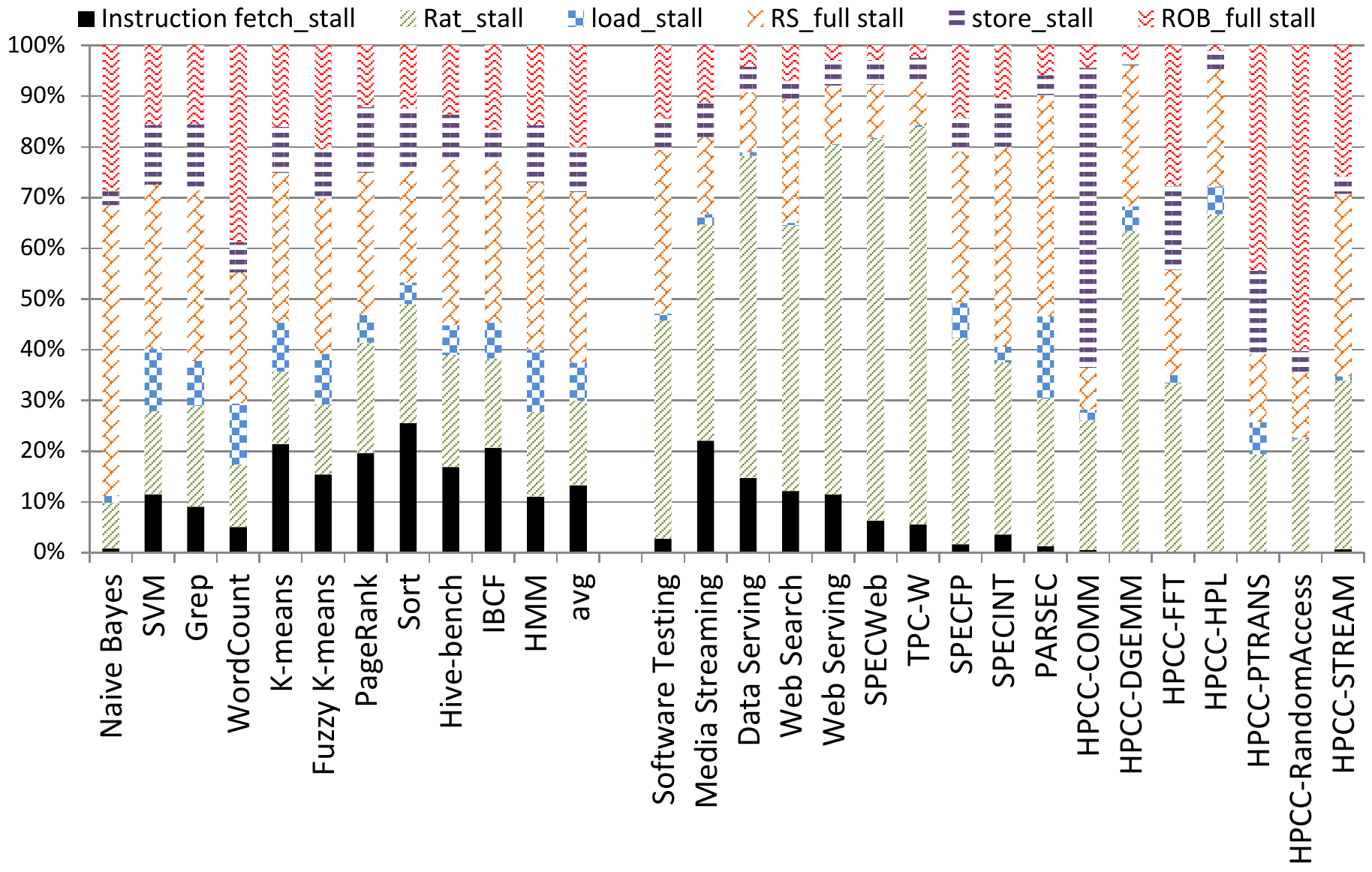}
\caption{Pipeline Stall Break Down of Each Workload}\label{stall}
\end{figure}

Figure~\ref{stall} presents those major stalls in pipelines for each workload including instruction fetch stalls,
RAT stalls, load buffer full stalls, store buffer full stalls,
RS full stalls, and ROB full stalls.
We can get the blocked cycles of those kind of stalls mentioned above by using hardware performance
counters.
Different kinds of pipeline stalls may occur simultaneously, that is to say, the stall
cycles may overlap.
For example,  when the back end is stalled due to \emph{RS} full, the front end can also be stalled due to L1 instruction cache misses.
So in Figure~\ref{stall}, we report the normalized values of the stalled cycles.
We calculate the normalized value by using the following way:
we sum up all the blocked cycles for all kinds of stalls as the total blocked cycles.
Then we divide each kind of stall's blocked cycles by the total blocked cycles as their percentage in Figure~\ref{stall}.

Different from HPCC, PARSEC and SPEC CPU2006 workloads,
the big data analytics workloads and service workloads suffer from notable instruction fetch stalls,
which are mainly caused by
L1 instruction cache miss, ITLB (Instruction Translation Lookaside Buffer)
miss or ITLB fault, reported in front-end performance data in Section~\ref{front}.
The notable instruction fetch stalls indicates the front end inefficiency. Our observation corroborates  the previous work~\cite{ferdman2011clearing}.
The front end inefficiency may caused by high-level languages, third-party libraries, deep software stacks used by the big data analytics and service workloads.
The most possible reason is that the complicated software stack and middle-ware increase the binary size of the whole application even though some of them only implement a simple algorithm.

We also find that there are notable differences in terms of stalls breakdown between the big data analytics workloads and the service workloads (including four service workloads of CloudSuite, \emph{SPECWeb} and \emph{TPC-W}).
The latter workloads own a large percentage of RAT stalls, which may be caused by partial register stalls or
register read port conflicts. While the big data analytics workloads suffer from more RS stalls and ROB stalls, which are caused by limited RS and ROB entries.
\emph{RAT} and instruction fetch stalls occur before instruction entering the out-of-order part of the pipeline while the RS and ROB stalls occur at the out of order part of the pipeline.
The service workloads (including \emph{Media Streaming, Data Severing, Web Severing,
Web Search, SPECweb and TPC-W}) have 63\% RAT
stalls and 12\% instruction fetch stalls on average, whereas the big data analytics workloads have about 37\% RS full stalls and 20\% ROB full stalls on average.
So we can find that the big data analytics workloads suffer more stalls in the out-of-order part of the pipeline, while the service workloads suffer more stalls in the in-order part of the pipeline.
Further investigation is necessary to understand the root cause behind the differences between two kinds of workloads.


For the HPCC workloads are composed of micro benchmarks and kernel programs, different programs focus on a specific aspect of the system. So their stall data vary dramatically from each other in Figure~\ref{stall}.


\emph{\textbf{Implications:}}

Corroborating previous work~\cite{ferdman2011clearing}, both the big data analytics workloads and the service workloads suffer from notable front-end stalls (i.e. instruction fetch stalls).
The instruction fetch stall means that the front end has to wait for fetching instructions,
which may be caused by two factors: deep memory hierarchy with long latency in modern processor~\cite{ferdman2011clearing}, and large binary size complicated by high-level language, third-party libraries and deep software stacks. And we verify that the software stack makes contribute to the front end stall for big data analytics workloads in Section~\ref{corr}.

However, we note the significant differences between the big data analytics workloads and the service workloads in terms of stall breakdown:
the big data analytics workloads suffer more stalls in the out-of-order
part of the pipeline,
while the service workloads suffer
more stalls before instructions entering the out-of-order part.
This observation can give us some implications
about how to alleviate the bottlenecks in pipeline,
although one well known consequence is that right after of alleviating the bottleneck, the next bottleneck emerges~\cite{thomadakis2011architecture}.


\subsection{Front-end Behaviors} \label{front}
\begin{figure}
\centering
\includegraphics[scale=0.7]{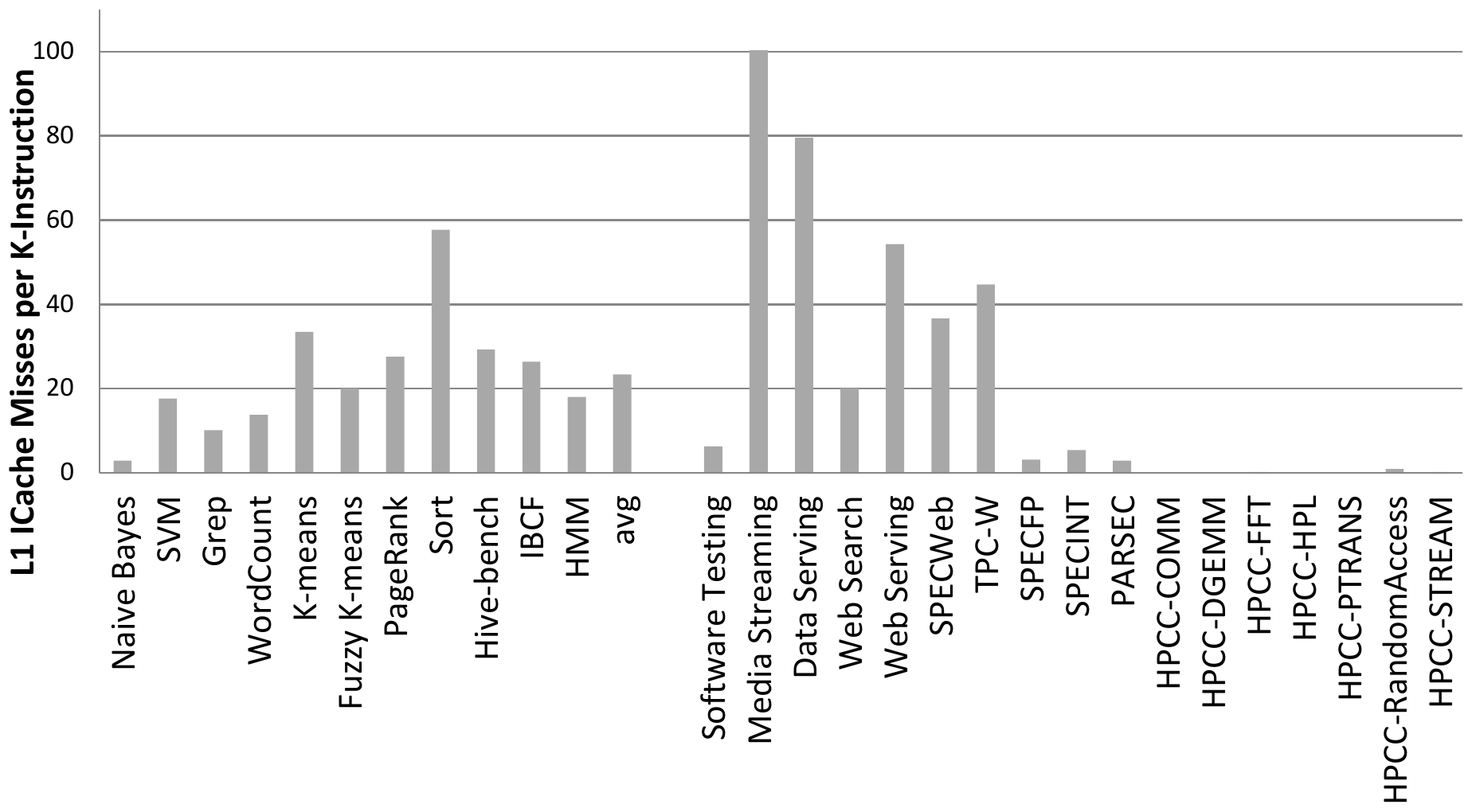}
\caption{L1 Instruction Cache misses per thousand instructions.}\label{L1cache}
\end{figure}

The instruction-fetch stall will prevent core from
making forward progress due to lack of instructions.
Instruction cache and instruction Translation Look-aside Buffer (TLB) 
are two fundamental components, which  must be accessed when fetching instructions from memory.
Instruction cache is the place where the fetch unit
directly get instructions.
TLB stores page table entries (PTE),
which are used to translate virtual addresses to physical addresses.
Each time a virtual memory access, the processor searches the
TLB for the virtual page number of the page that is being accessed.
If a TLB entry is found with a matching virtual page
number, a TLB hit occurs and the processor can use the retrieved physical address
to access memory.
Otherwise there is a TLB miss, the processor has to look up the page table, which called a page walk.
The page walk is an expensive operation.
With a three-level page table, three memory
accesses would be required. In other words, it would result in
four physical memory accesses.

\begin{figure}
\centering
\includegraphics[scale=0.7]{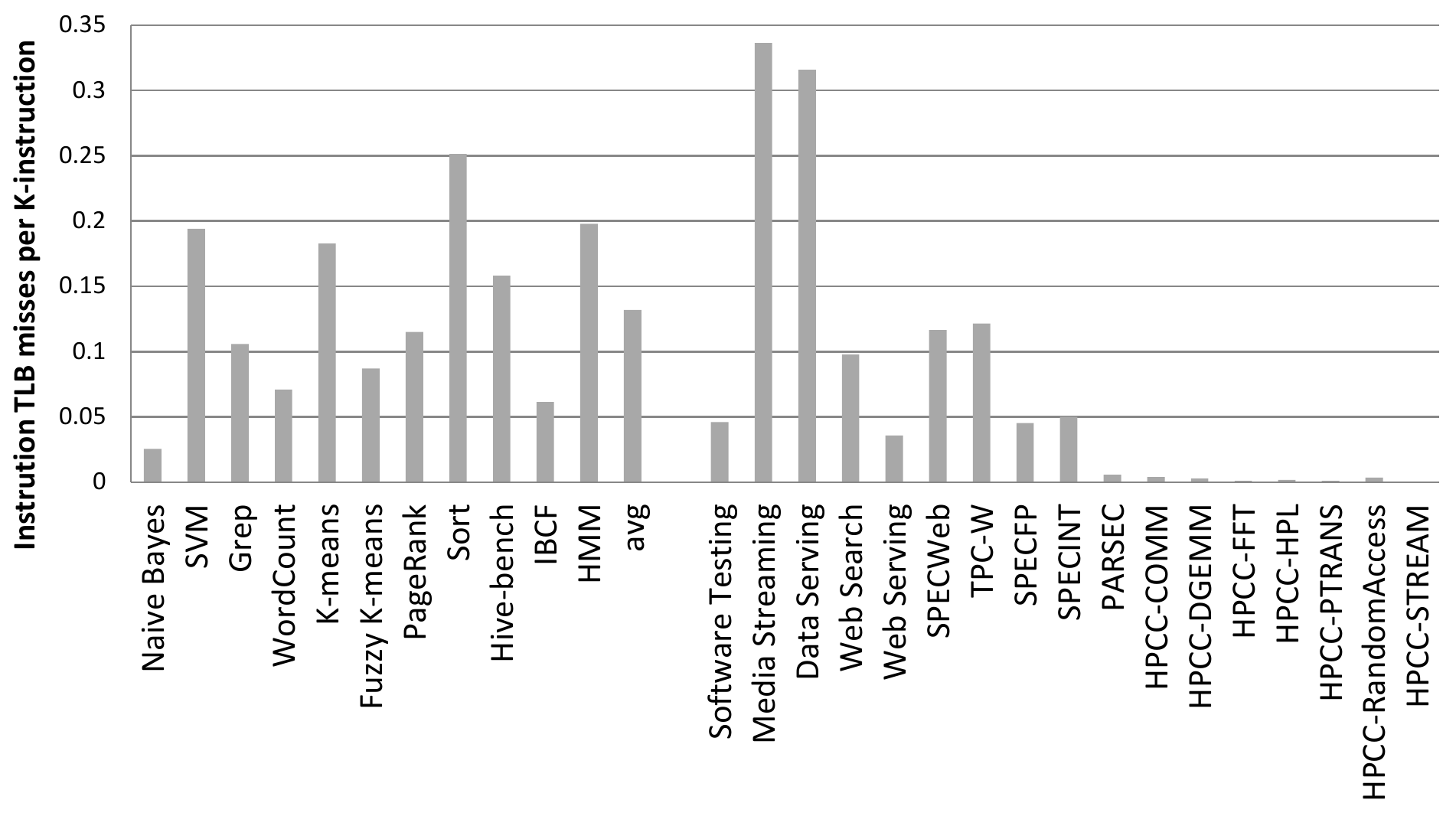}
\caption{Instruction TLB misses per thousand instructions.}\label{itlb}
\end{figure}

Figure~\ref{L1cache} and Figure~\ref{itlb} present the L1 instruction cache misses and the instruction TLB misses, which trigger page walks, per thousand instructions, respectively.
On average, the big data analytics workloads generate about 23 L1
instruction cache misses per thousand instructions.
They own higher L1 instruction cache misses
than those of  \emph{SPECINT}, \emph{SPECFP}, and all the HPCC workloads.
Most of the big data analytics applications have less L1 instruction cache misses than those of the service workloads including \emph{Media Streaming, Data Severing, Web Serving, TPC-W} and \emph{SPECweb}.
\emph{Media streaming} has a larger instruction footprint and suffers from
severe L1 instruction cache misses, whose L1 instruction cache misses are
about three times more than the average of that of the big data analytics workloads.
Higher L1 instruction cache misses result in higher instruction fetch stalls as shown in
Figure~\ref{stall}, indicating less efficiency of the front-end.
For most of the others benchmarks, the L1 instruction cache misses are really very rare, especially the
HPCC workloads, whose instruction footprint is relatively small.


Consistent with the performance trend of L1 instruction cache misses, the big data analytics workloads' instruction TLB misses are more frequently than those of \emph{SPECINT}, \emph{SPECFP, PARSEC}, and all HPCC workloads.
Some service workloads (\emph{Media Streaming} and \emph{Data Serving} workloads) have more
instruction TLB misses than those of the big data analytics workloads.
Page walks will cause a long latency instruction fetch stall, waiting for correct physical addresses so as to fetch instructions, and  hence result in inefficiency of front end.
Among the big data analytics workloads, \emph{Naive Bayes} is an exception  with the fewest L1 instruction cache misses and instruction TLB misses, so it is not enough to represent the spectrum of all big data analytics workloads.


\emph{\textbf{Implications:}}

Improving the L1 instruction cache and instruction TLB performance can improve the performance of  data center workloads, especially the service workloads.
The third-party libraries and software stacks used by data center workloads may enlarge the
binary size of applications and further aggravate the inefficiency of instruction cache and TLB.
So when writing the program (with the support of third-party libraries and software stack), the engineers should pay more attention to the code size and the potential burden to pipeline front end.


\subsection{Unified Cache and Data TLB Behaviors} \label{mhb}
The manufacturers of processors introduce a deep memory hierarchy to reduce the
performance impacts of memory wall.
Nearly all of the modern processors own three-level caches.
A miss  penalty of last-level cache can reach
up to several hundred cycles in modern processors.

\begin{figure}
\centering
\includegraphics[scale=0.67]{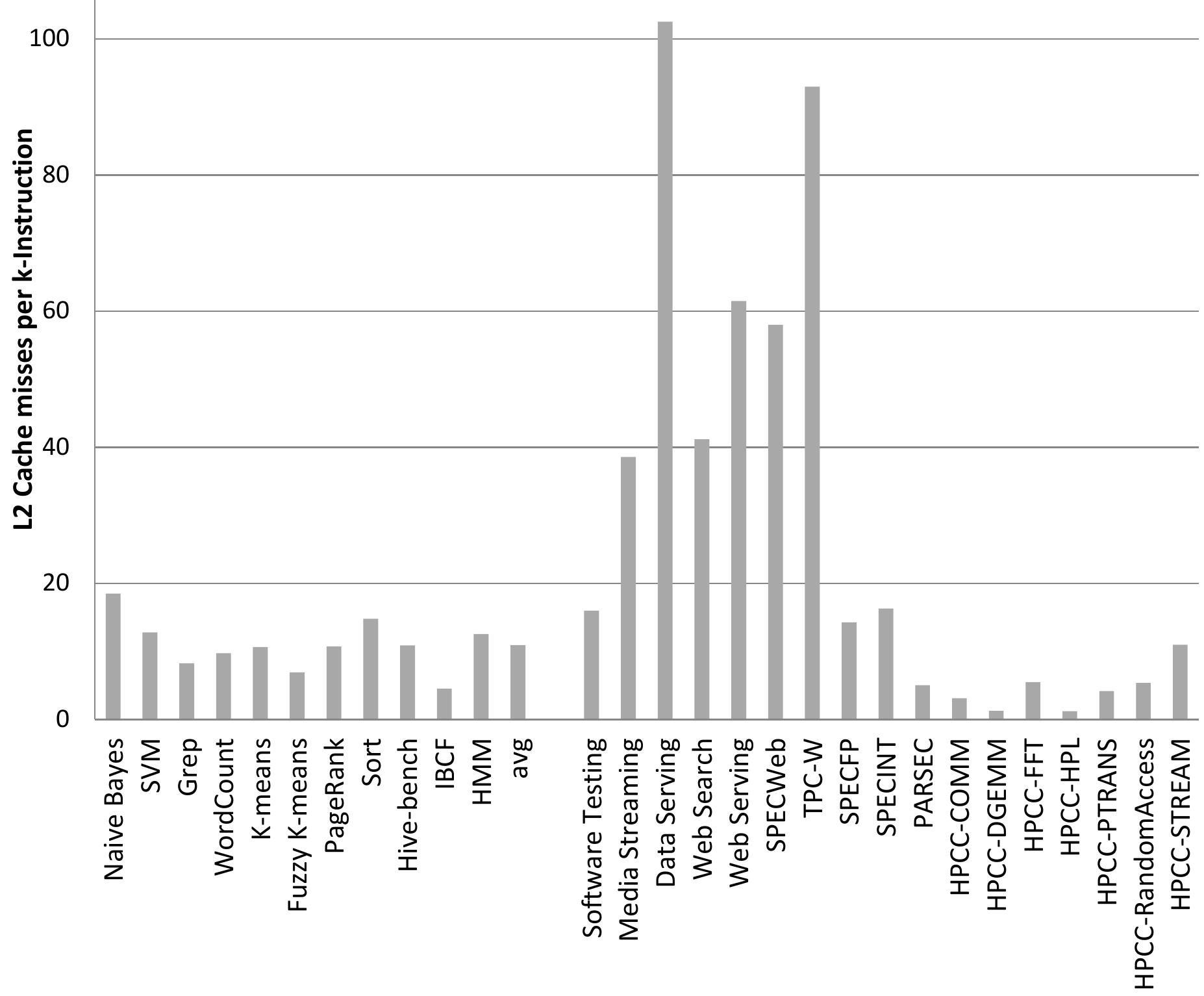}
\caption{L2 cache misses per thousand instructions.}\label{cache_miss}
\end{figure}

\begin{figure}
\centering
\includegraphics[scale=0.7]{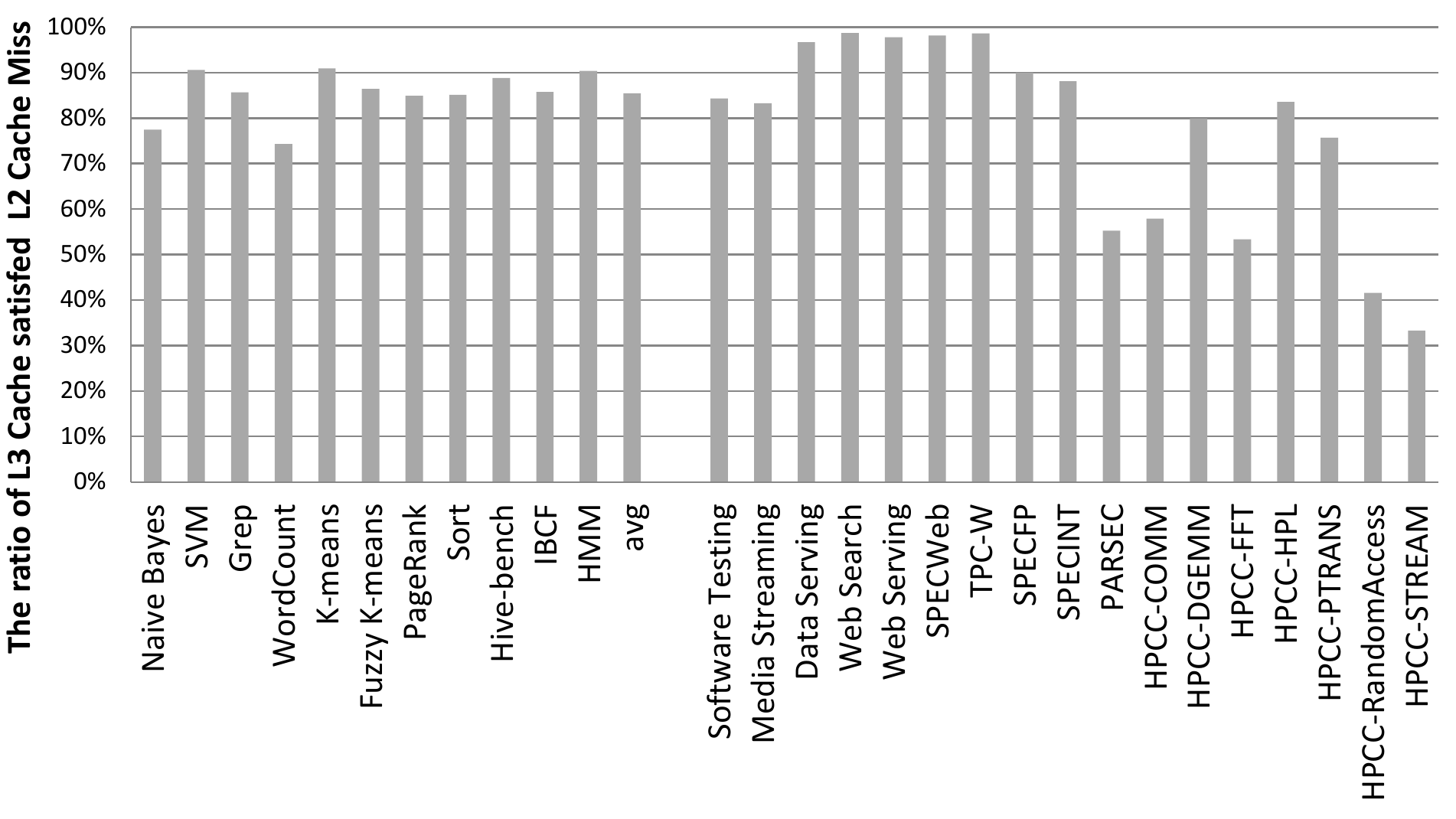}
\caption{The ratio of L3 cache satisfying L2 cache misses.}\label{l3cachel2}
\end{figure}

Figure~\ref{cache_miss} shows the L2 cache MPKI (misses
per thousand instructions).  Figure~\ref{l3cachel2} reports the ratio of L3 cache hits over L2 cache misses.
This ratio can be calculated by using  Equation~\ref{eq:equ1}.
Please note that we do not analyze the L1 data cache statistics for the miss penalty can be hidden by
the out-of-order cores~\cite{karkhanis2004first}.    

\begin{equation} \label{eq:equ1}
  ratio =\frac{L2 ~ cache ~misses- L3 ~cache ~misses}{L2 ~cache~ misses}
\end{equation}

For most of the big data analytics workloads, they have lower L2 cache misses (about 11 L2 cache MPKI on average) than those of the service workloads (about 66 L2 cache MPKI on average) while higher than those of the HPCC workloads.
The L2 cache statistic indicates the big data analytics workloads own better locality than the service workloads.
The HPCC workloads have different localities as the official web site mentioned, which can explain
the different cache behaviors among the HPCC workloads. 

From Figure~\ref{l3cachel2}, we can find that for both the big data analytics workloads and service workloads, the average ratio of L2 cache misses that are hit in L3 cache (85.5\% for the big data analytics workloads and 95.5\% for the service workloads) is higher than that of the PARSEC and HPCC workloads. We can conclude that for most of the big data analytics and service workloads, modern processor's LLC is large enough to cache most of data missed from L2 cache.



\begin{figure}
\centering
\includegraphics[scale=0.7]{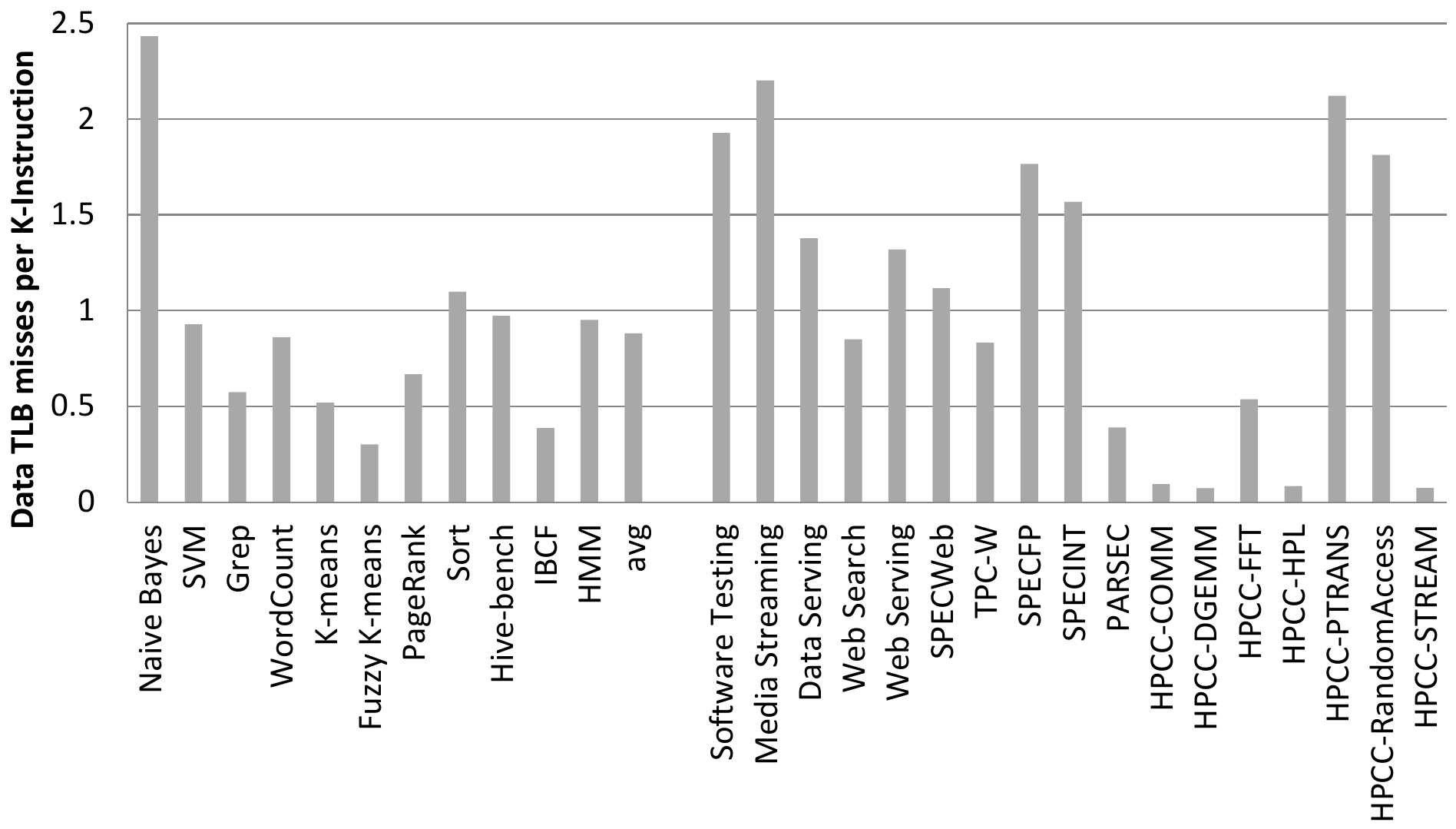}
\caption{DTLB Misses per Thousand Instructions Retired.}\label{dtlb}
\end{figure}

Figure~\ref{dtlb} shows the data TLB misses per thousand instructions.
For most of the big data analytics workloads with the exception of \emph{Naive Bayes}, the data TLB misses are less than most of the service workloads and SPEC CPU2006 workloads (\emph{SPECINT} and \emph{SPECFP}), but higher than most of the HPCC workloads with the exception of \emph{HPCC-RandomAcess} and \emph{HPCC-PTRANS}.
That means the data locality of most of the big data analytics workloads is much better than that of the service workloads.



\emph{\textbf{Implications:}}

For the big data analytics workloads, L2 cache is acceptably effective when compared with service workloads. They have lower
L2 cache MPKI than that of the service workloads, while higher than that of the HPCC workloads.
Meanwhile, for the big data analytics and service workloads, most of L2 cache misses are hit in L3 cache, indicating L3 cache is pretty effective.
Modern processors dedicate approximately half of the die
area for caches, and hence optimizing the LLC capacity properly may not only reduce the memory access latency but also improve the energy-efficiency of processor and save the die area.
For the service workloads, our observation corroborate the previous work~\cite{ferdman2011clearing}: the L2 cache is ineffective.
\subsection{Branch Prediction}

The branch instruction prediction accuracy
is one of the most important factor that directly affects the
performance.
Modern out-of-order processors introduce a functional unit (e.g. Branch Target Buffer)
to predict the next branch to avoid pipeline stalls due to branches.
If the predict is correct, the pipeline will continue. However,
if a branch instruction is mispredicted,
the pipeline must flush the wrong instructions and fetch the correct ones, which will cause at least a dozen of cycles' penalty.
So  branch prediction is not a trivial issue in the pipeline.

\begin{figure}
\centering
\includegraphics[scale=0.7]{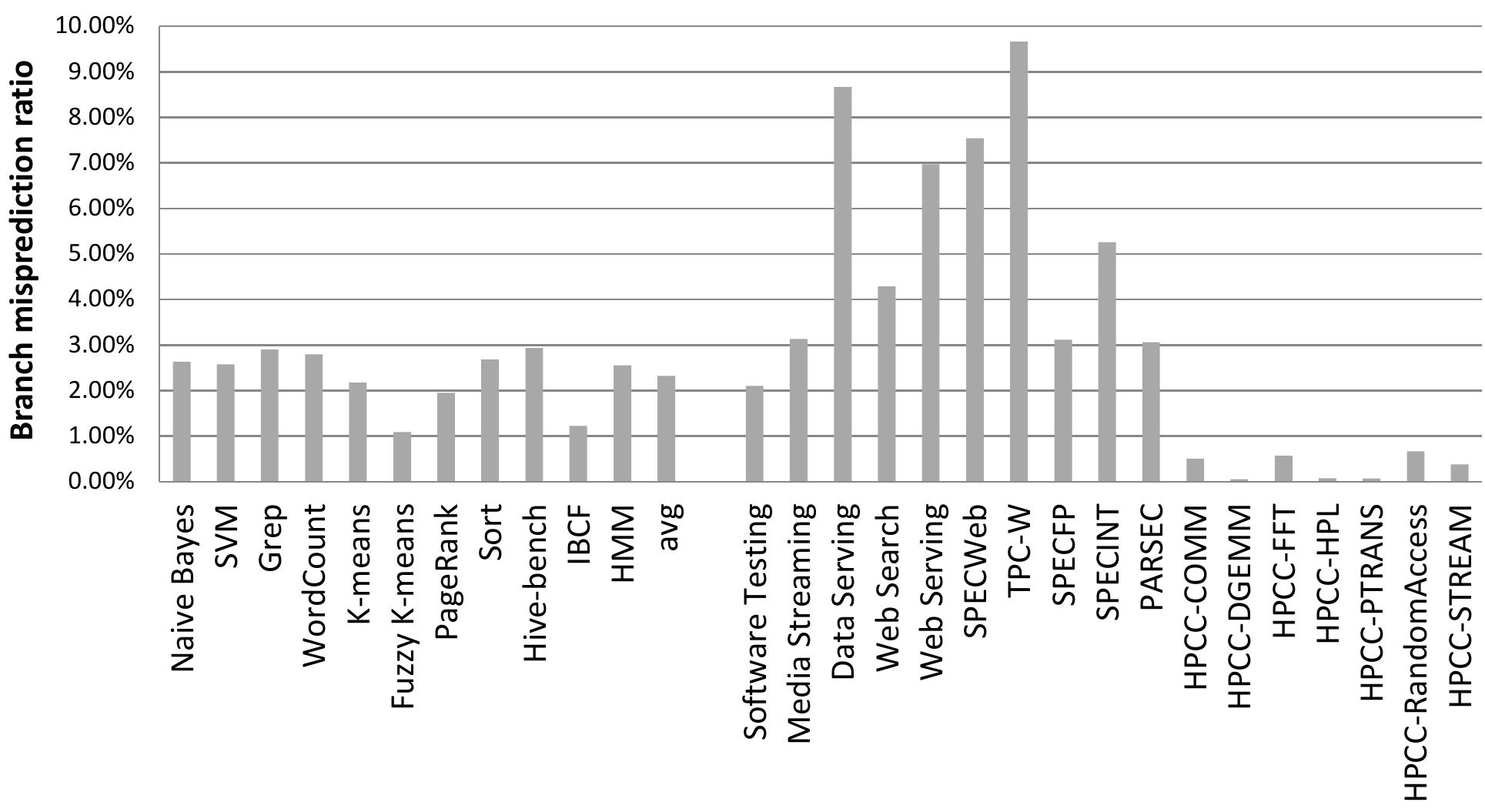}
\caption{Branch Miss-prediction ratio.}\label{branch-miss}
\end{figure}

Figure \ref{branch-miss} presents the branch miss prediction ratios of each workload.
We find that most of the big data analytics workloads own a lower branch misprediction ratio in comparison with that of the service workloads and SPEC CPU workloads.
The HPCC workloads own very low misprediction ratios because the branch
logic codes of the seven micro
benchmarks are simple and the branch behaviors have great regularity.
The low misprediction ratios of the big data analytics workloads indicates that most of the branch
instructions in the big data analytics workloads have simple patterns. 
The simple patterns are conducive to BTB (Branch Target Buffer) to predict whether the next
branch needs to jump or not.
For  the big data analytics workloads,  simple algorithms
chosen for big data always beat better
sophisticated algorithms~\cite{rajaraman2008more}, which may be the possible  reason for their  low misprediction ratios.

\emph{\textbf{Implications: }}

Modern processors invest heavily in silicon real estate and algorithms for the branch
prediction unit in order to minimize the frequency and the impact of wrong branch prediction.
For the big data analytics workload, the misprediction ratio is lower than most of the compared workloads,
even for the CPU benchmark --- SPECINT, which implies that the branch predictor of modern processor
is good enough for the big data analytics workloads. A simpler branch predictor may be preferred so
as to save power and die area.

\section{Correlation Analysis} \label{corr}
Correlation analysis can measures the relationship between two items and show the statistical relationships involving dependence~\cite{stigler1989francis}.
Correlations are useful because they can indicate a predictive relationship that can be exploited in practice.
The Pearson's correlation coefficient is the most popular method for correlation analysis.
It is a measure of the linear correlation between two variables. It is defined as the covariance of the two variables divided by the product of their standard deviations,
which is represented by Equation~\ref{equ1}.

\begin{equation} \label{equ1}
\begin{aligned}
\rho(X,Y)=corr(X,Y)=\frac{cov(X,Y)}{\sigma_X \sigma_Y}\\
=\frac{E[(X-\mu_X)(Y-\mu_Y)]}{\sigma_X \sigma_Y}
\end{aligned}
\end{equation}

The Pearson's correlation coefficient ranges from -1 to 1. The absolute value of the correlation coefficient shows the dependency. The bigger the absolute value, the stronger the correlation between the two variables. The positive number means positive correlation, and vice versa.
A value of 1 implies that a linear equation describes the relationship between X and Y perfectly. 
A value of 0 implies that there is no linear correlation between the variables.

CPI (Cycles Per Instruction) refers to the number of processor cycles an instruction consumed in the pipeline. The more cycles a processor takes to complete an instruction, the poorer the performance of the application in the pipeline.
So CPI is the metric used to evaluate the application's performance on the pipeline from the perspective of micro-architecture.
In order to decide which factors affect the CPI performance,
we compute the correlation coefficients of the above micro-architecture level characteristics in Section~\ref{characterization_BDbenchmarks} with CPI.
We use CPI to perform correlation analysis,
because most of the metrics shown in Section~\ref{characterization_BDbenchmarks} will potentially increase the CPI value, so we can see the positive correlations between CPI and other metrics.
We analyze workloads' instruction level parallelism with IPC, which is the multiplicative inverse of CPI, in Section~\ref{sub_ipc}, and use CPI to represent processor performance in this section.

Tables~\ref{DAco} and \ref{otherco} show the correlation coefficients of each
of the above characteristics with CPI.
In those Tables, we only present the first five metrics that own the highest correlation coefficients
with CPI.
The metrics and the corresponding correlation coefficients are shown in those table in a decreasing order. From Tables~\ref{DAco} and \ref{otherco}, we can make the following observations.

In contrast to traditional workloads and service workloads, most of big data analytics workloads' CPI performance is sensitive to load buffer performance, i.e. the load buffer full stall has a strong positive correlation with CPI.
For the service workloads, the metrics that affect the workloads' CPI are more diverse. Some are very sensitive to instruction fetch stall, such as  \emph{Media Streaming}. Some are sensitive to branch instruction execution situation including branch misprediction ratio and branch instruction ratio.
Different from all of other workloads, the chip multiprocessors (PARSEC) and high performance (HPCC) workloads are more sensitive to L3 cache performance.
We also can find that
nearly for all workloads, the L2 cache performance and kernel-mode instruction have positive correlations with CPI.
Most of the correlation coefficients between L2 cache miss ratio (or kernel-mode instructions) and CPI are no less than 0.6, which indicates strong positive correlations.
Both instruction and data TLB performance also have impacts on CPI performance for nearly all workloads we investigated, because of the large miss penalty the TLB miss owns~\footnote{Our Xeon processor have a tow-level TLB. The first level has separate
instruction and data TLBs. The second level is shared. We measure a TLB miss at both level, which means a page walk happened.}.

\emph{\textbf{Implications: }}

Although big data analytics workloads own notable instruction fetch stalls, the instruction fetch stall is not the factor that affects the CPI performance most.
We can find that the instruction fetch stall does not appear in Table~\ref{DAco} with high correlation coefficient value
for most of big data analytics workloads. The only exceptions are \emph{Naive Bayes} and \emph{Sort}, however the correlation coefficients are small, only 0.198 and 0.424 respectively, which indicates  very weak correlations.
The instruction fetch stall play a very critical role in program performance by preventing the pipeline from making forward; however it is not the optimization point with highest priority for big data analytics applications.
There are many potential optimization points in modern superscalar processors as previous work found e.g., on-chip bandwidth, die area and etc~\cite{ferdman2011clearing}.
According to our correlation analysis in this section,
architects should focus on improving TLB performance and the private unified cache (L2 cache for our processor) performance with the highest priority for big data analytics workloads. Just as a page walk, which is caused by a TLB miss, is a very expensive operation. Optimizations should focus on reducing the miss penalty either by enlarging the TLB capacity to hold more entries or by accelerating the speed that refills the TLB.
For the private unified cache (L2 cache), the big data analytics workloads have pretty good performance from the perspective of cache miss ratio.
So the miss penalty of private unified cache should be the optimization point for big data analytics workloads.
Considering that the last level cache can hold most of the misses from previous cache levels as mentioned in Section~\ref{mhb}, reducing the capacity of last level cache appropriately may be a good choice, just as we suggested in Section~\ref{mhb}.
A smaller last level cache can not only reduce the L2 cache miss penalty but also improve the energy efficiency and save die area. However for chip multiprocessors (PARSEC) and high performance (HPCC) workloads, reducing the last level cache capacity may not be a good choice since their performance is very sensitive to L3 cache miss ratio.



\begin{table*}
\caption{Big data analytics and service workloads correlation coefficients}\label{DAco}
\center \sffamily \scriptsize
\begin{tabular}{|c|c|c|c|c|c|} \hline
 Workload& \multicolumn{2}{c|}{Correlation Coefficients} & Workload& \multicolumn{2}{c|}{Correlation Coefficients} \\ \hline
 \multirow{5}{*}{Naive Bayes} & L3 cache miss & 0.631 & \multirow{5}{*}{SVM} &L2 cache miss &0.949 \\ \cline{2-3} \cline{5-6}
 & L2 cache miss	&0.419 &  &Instruction TLB miss &0.930 \\ \cline{2-3} \cline{5-6}
 & Data TLB miss & 0.334 & &Data TLB miss &0.901 \\ \cline{2-3} \cline{5-6}
 & L1 Instruction cache miss	& 0.237 & & Kernel-mode instruction &0.890\\ \cline{2-3} \cline{5-6}
 &Instruction fetch stall &	0.198 & & Load buffer full stall& 0.875 \\ \hline
\multirow{5}{*}{Grep} &L2 cache miss & 0.967 & \multirow{5}{*}{WordCount}  &Reservation station full stall&0.763\\ \cline{2-3} \cline{5-6}
&Instruction TLB miss	&0.944 & &L2 cache miss & 0.626\\ \cline{2-3}  \cline{5-6}
&Data TLB miss	&0.893 & &Instruction TLB miss	&0.618 \\ \cline{2-3} \cline{5-6}
&Kernel-mode instruction &0.834 & &Data TLB miss	&0.571\\ \cline{2-3} \cline{5-6}
&Load buffer full stall	&0.728 & &Branch misprediction	&0.501\\ \hline
\multirow{5}{*}{K-means} & Data TLB miss &0.954 & \multirow{5}{*}{Fuzzy K-means} & L2 cache miss	 &0.918 \\ \cline{2-3} \cline{5-6}
&Instruction TLB miss &0.914 & &Instruction TLB miss&	0.905\\ \cline{2-3} \cline{5-6}
&Load buffer full stall	&0.822 & &Data TLB miss	&0.895\\ \cline{2-3} \cline{5-6}
&L2 cache miss	&0.821& &Kernel-mode instruction &0.837\\ \cline{2-3} \cline{5-6}
&Kernel-mode instruction &0.816 & &Load buffer full stall&	0.748\\ \hline
\multirow{5}{*}{PageRank}& Data TLB miss &0.872 & \multirow{5}{*}{Sort} &Load buffer full stall& 0.592\\ \cline{2-3} \cline{5-6}
& Kernel-mode instruction 	&0.743 & & Data TLB miss &	0.493\\ \cline{2-3} \cline{5-6}
&L2 cache miss	&0.679 & &L2 cache miss	&0.485\\ \cline{2-3} \cline{5-6}
&Instruction TLB miss	&0.586 & &Instruction fetch	stall&0.424\\ \cline{2-3} \cline{5-6}
&L1 instruction cache miss&0.405 & &Kernel-mode instruction	&0.423\\ \hline
\multirow{5}{*}{Hive Bench} &L2 cache miss	& 0.901& \multirow{5}{*}{IBCF} &L2 cache miss	 &0.809\\ \cline{2-3}  \cline{5-6}
&Data TLB miss	&0.856& &Data TLB miss &	0.793\\ \cline{2-3}  \cline{5-6}
&Reservation Station stall	 &0.815& &Kernel-mode instruction &	0.607\\ \cline{2-3}  \cline{5-6}
&Kernel-mode instruction	&0.766& &L3 cache miss	& 0.478\\ \cline{2-3}  \cline{5-6}
&Instruction TLB miss	&0.555& &Branch misprediction	& 0.478\\ \cline{1-3} \Xcline{4-6}{1.2bp}
\hline
\multirow{5}{*}{HMM} & Data TLB miss	&0.894  & \multirow{5}{*}{TPC-W} & Data TLB miss	 &0.972 \\ \cline{2-3} \cline{5-6}
&L2 cache miss	&0.874 &&L2 cache miss	&0.939\\ \cline{2-3} \cline{5-6}
&Instruction TLB miss	&0.783 &&Instruction TLB miss	&0.855\\ \cline{2-3}\cline{5-6}
&Branch misprediction 	&0.612 &&Branch instruction retired	& 0.577\\ \cline{2-3} \cline{5-6}
&Kernel-mode	instruction& 0.609 &&Kernel-mode	instruction & 0.557\\ \Xcline{1-3}{1.2bp} \cline{4-6}
\hline
\multirow{5}{*}{Software Testing} & Instruction TLB miss	&0.983 &\multirow{5}{*}{Media Streaming}& Branch instruction retired&	0.954 \\ \cline{2-3} \cline{5-6}
& L2 cache miss	& 0.978 & &L1 instruction miss	&0.902 \\ \cline{2-3} \cline{5-6}
& L3 cache miss	& 0.977 & &Instruction fetch stall &0.872\\ \cline{2-3} \cline{5-6}
&Data TLB miss	&0.886 &&Kernel-mode instruction	&0.821\\ \cline{2-3} \cline{5-6}
&Instruction fetch stall &0.877 & &ReOrder Buffer stall	&0.808 \\ \hline
\multirow{5}{*}{Data Serving} &Data TLB miss	&0.971& \multirow{5}{*}{Web Search} &Data TLB miss & 0.995\\ \cline{2-3} \cline{5-6}
&L2 cache miss &0.953& &L2 cache miss& 0.994 \\ \cline{2-3}\cline{5-6}
& Instruction TLB miss	& 0.948 & &Instruction TLB miss&0.988\\ \cline{2-3}\cline{5-6}
&Kernel-mode	instruction &0.925 &&Kernel-mode instruction&0.963\\ \cline{2-3}\cline{5-6}
& Branch misprediction	& 0.925& &Load buffer full stall&0.912 \\ \hline
\multirow{5}{*}{Web Serving} & Data TLB miss & 0.981 & \multirow{5}{*}{SPECWeb} &Instruction TLB miss &	 0.920\\ \cline{2-3}\cline{5-6}
& L2 cache miss	 & 0.947 & &Data TLB miss	&0.845 \\ \cline{2-3}\cline{5-6}
&Instruction TLB miss	&0.911 & &Kernel-mode instruction &0.786 \\ \cline{2-3}\cline{5-6}
&Reservation Station full stall &	0.566& &Store buffer full stall &0.641\\ \cline{2-3}\cline{5-6}
&Kernel-mode instruction &	0.545& &L1 instruction cache miss	& 0.592 \\ \hline
\end{tabular}
\end{table*}

\begin{table}
\caption{Traditional workloads's correlation coefficients}\label{otherco}
\center \scriptsize \sffamily
\begin{tabular}{|c|c|c|} \hline
Workload& \multicolumn{2}{c|}{Correlation Coefficients} \\ \hline
\multirow{5}{*}{PARSEC} &L3 cache miss & 0.899\\ \cline{2-3}
 &L2 cache miss	&0.513\\ \cline{2-3}
 &Data TLB miss	&0.375\\ \cline{2-3}
 &L1 instruction miss	& 0.295\\ \cline{2-3}
&Instruction fetch stall &0.877 \\ \hline
\multirow{5}{*}{HPCC-COMM} &L3 cache miss& 0.999 \\ \cline{2-3}
 &L2 cache miss& 0.988 \\ \cline{2-3}
&Kernel-mode	instruction &0.980 \\ \cline{2-3}
 & Branch misprediction	&0.495\\ \cline{2-3}
 &L1 instruction miss	&0.400\\ \hline
\multirow{5}{*}{HPCC-DGEMM} & L3 cache miss	 &0.988\\ \cline{2-3}
 &L2 cache miss	&0.751 \\ \cline{2-3}
&Data TLB miss	&0.662\\ \cline{2-3}
 &Branch misprediction	&0.611\\ \cline{2-3}
 &Instruction TLB miss	&0.353 \\ \hline
\multirow{5}{*}{HPCC-FFT} &L3 cache miss	 &0.997 \\ \cline{2-3}
& L2 cache miss	&0.961 \\ \cline{2-3}
& Instruction TLB miss	&0.828\\ \cline{2-3}
& L1 instruction miss	&0.459\\ \cline{2-3}
&  kernel-mode instruction&0.435
\\ \hline
 \multirow{5}{*}{HPCC-HPL} & L3 cache miss&	 0.859 \\ \cline{2-3}
 & L2 cache miss	&0.834 \\ \cline{2-3}
  &Kernel-mode instruction	&0.597 \\ \cline{2-3}
 &Data TLB miss	&0.589\\ \cline{2-3}
 &Instruction TLB miss	&0.435\\ \hline
 \multirow{5}{*}{HPCC-PTRANS} & L3 cache miss & 0.871\\ \cline{2-3}
 &Kernel-mode instruction	& 0.812 \\ \cline{2-3}
 &Instruction TLB miss	& 0.809\\ \cline{2-3}
&Reservation buffer full store	&0.734\\ \cline{2-3}
&ReOder Buffer full stall	& 0.645\\ \hline
\multirow{5}{*}{HPCC-Random Access} &L3 cache miss &0.999\\ \cline{2-3}
&L2 cache miss &0.999\\ \cline{2-3}
&Data TLB miss	&0.999\\ \cline{2-3}
&L1 instruction miss	&0.911\\ \cline{2-3}
&Instruction fetch	stall&0.890\\ \hline
\multirow{5}{*}{HPCC-Stream} &L3 cache miss	&0.995\\ \cline{2-3}
& L2 cache miss	&0.978 \\ \cline{2-3}
&L1 instruction cache miss	&0.873\\ \cline{2-3}
& Data TLB miss	&0.674 \\ \cline{2-3}
& Application instruction retired  & 0.398\\ \hline
\multirow{5}{*}{SPEC INT} &L2 cache miss	& 0.767 \\ \cline{2-3}
&Data TLB miss	&0.699 \\ \cline{2-3}
&Instruction TLB miss	&0.493\\ \cline{2-3}
&Kernel-mode instruction	&0.454 \\ \cline{2-3}
&L3 cache miss	&0.389\\ \cline{1-3}
\multirow{5}{*}{SPEC CFP} &Data TLB miss &0.719\\ \cline{2-3}
&Instruction TLB miss	&0.582\\ \cline{2-3}
&Kernel-mode instruction	&0.549\\ \cline{2-3}
&L2 cache miss	&0.452\\ \cline{2-3}
&L3 cache miss	& 0.308\\ \hline
\end{tabular}
\end{table}

\section{Software Stack Impact} \label{software}
Software stacks are being proposed to facilitate the development of big data analytics applications.
Those software stacks, such as Spark~\cite{zaharia2012resilient} and Hadoop~\cite{white2009hadoop}, have attracted a large number of users and companies in a short period of time~\cite{HadoopUsageReport,sparkpowerby}.
On one hand, the big data analytics software stack facilitates the programmer to write a big data analytics application without considering the messy details of data partitioning, task distribution, load balancing, failure handling and other warehouse-scale system details~\cite{ren2013hadoop,jia2014characterizing}.
On the other hand, the big data analytics software stack may affect the application behaviors for the software stack increase the call hierarchy of big data analytics applications.
Since all the big data analytics workloads we characterized in this paper are based on Hadoop software stack, we would like to investigate the Hadoop's impacts on modern
processors as a case study and show what programmers and architects
can learn from those impacts.

Hadoop has three different operation modes:  standalone (local) mode, pseudo-distributed mode and fully distributed mode~\cite{lam2010hadoop}. In the standalone mode, the Hadoop will run completely on the local machine. It does not use HDFS, nor will it launch any of the Hadoop daemons. The pseudo-distributed mode is running Hadoop in a ``cluster of one'' with all daemons running on a single machine. And the fully distributed mode provides a production environment, which can manage a large number of nodes. All the big data analytics workloads are running in the fully distributed mode in previous sections.

We can take a glimpse at the big data software stack's impacts by comparing the application behaviors between standalone mode and (pseudo or fully) distributed mode.
Actually the standalone mode does not eliminate the impact of the software stack completely, but it eliminates the HDFS and daemon processes' impacts and further alleviates the software stack's impacts largely.
The standalone mode really provide us the chance that executes the same user application code but with less call hierarchy.

In this section, we chose the pseudo-distributed mode as the compared  running mode, which invoke the full software stack, for the pseudo-distributed mode eliminates the network factor brought by fully distributed mode. Table~\ref{hadoop_mode} shows the call hierarchy for those two modes.


\begin{table}
\caption{Hadoop call hierarchy among different modes.``Y'' means the corresponding mode will invoke the item. ``N'' means the corresponding mode will not invoke the item}\label{hadoop_mode}
\center \scriptsize \sffamily
\begin{tabular}{|c|c|c|} \hline
& Standalone&  Pseudo-distributed \\ \hline
Hadoop daemons &N& Y\\ \hline
HDFS & N & Y \\ \hline
MapReduce API & Y &Y\\ \hline
JVM & Y& Y\\ \hline
\end{tabular}
\end{table}

We choose eight applications to investigate the impact of the typical big data analytics software stack, i.e. Hadoop,  because the other three applications use third party libraries heavily~\footnote{HMM invokes ICTCLASS~\cite{ICTCLASS}; SVM invokes LIBSVM~\cite{chang2011libsvm} and Hive-bench invokes Hive.}, which may make it difficult to analyze the Hadoop software stack's impact.
For all the eight applications are running on a single node, we must drive them with smaller data sets to avoid overload. We use about 10 GB data set for each workload and use the same data set to drive applications running in different operation mode in order to eliminate the input data set factor.
We run the same application on different modes and collect the micro-architecture level metrics.

In the rest parts of this section, we mainly focus on investigating the software stack's impacts on the following aspects.
1) The instruction fetch unit performance. For the instruction fetch stall is a notable feature that differentiate big data analytics workloads from most of traditional workloads. We want to verify the impacts that the software stack has on instruction fetch unit.
2) The private unified cache, TLB and load buffer performance. Not only for they are the critical units along the data path, but also for they are the metrics that have strong correlations with CPI performance as elaborating in Section~\ref{corr}.
We want to know how the software stack affect those units' performance.
3) Kernel-mode instruction ratio. For it is also one of the metrics that has a strong correlation with CPI performance. In addition, we want to investigate on which level
do software stack introduced instructions executed.


\subsection{Instruction Fetch}
\begin{figure}
\centering
\includegraphics[scale=0.9]{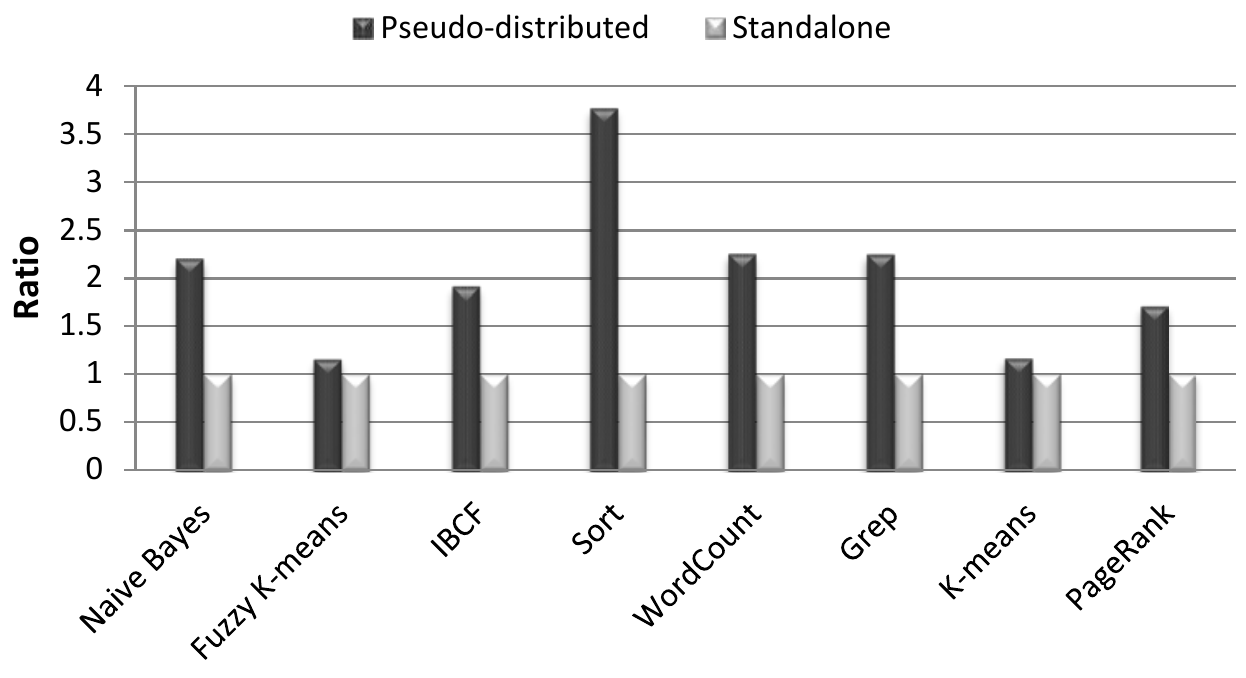}
\caption{Instruction Fetch Stall.}\label{ins-fetch}
\end{figure}

Figure~\ref{ins-fetch} shows the instruction fetch stall per cycle. It presents the normalized pseudo-distributed mode values using the standalone mode data as the baseline.
We can find that the software stack really has impacts on the instruction fetch unit. For all the workloads the pseudo-distributed mode has more instruction fetch stalls than that of standalone mode.
The ratio ranges from 1.17 to 3.77 and on average the pseudo-distributed workloads' instruction fetch stalls are 2.05 times of those of their standalone counterparts.
This implies that the software stack puts more pressure on the instruction fetch unit.
This phenomenon most probably caused by the increased application binary size. 
With the participation of full software stack,
a large number of instructions are needed to implement the strategies and mechanisms
provided by software stack such as fault tolerant, which increases big data application's binary size and aggravates the inefficiency of instruction fetch unit.


\subsection{Data Path performance}

\begin{figure}
\centering
\includegraphics[scale=0.9]{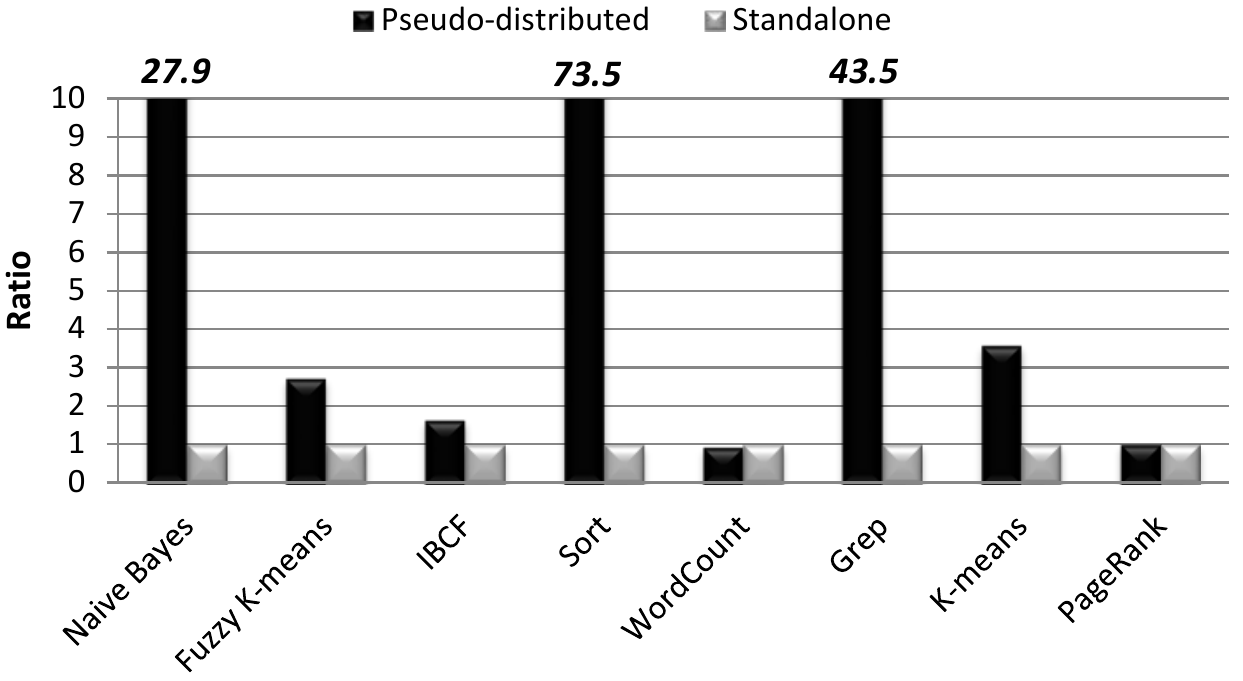}
\caption{Load buffer full stall.}\label{local-load}
\end{figure}

Load buffer is another very important unit along data path, and it also has great impact on CPI performance for big data analytics workloads as discussed in Section~\ref{corr}.
Each load micro-operation requires a load buffer entry and will access the data TLB and data cache.
The load buffer keeps track of in-flight loads in the out of order processor.
The load buffer full stall event records the cycles of stall due to lack of load buffer entry.
We calculate the ratio of load buffer full stall cycles to the total cycles the application used.
We also give the normalized value in Figure~\ref{local-load} by using the standalone mode workloads as baseline.
From Figure~\ref{local-load}, we can find that most of the applications running at pseudo-distributed mode have more load buffer full stalls than their standalone counterpart.  The only exceptions are \emph{WordCount} and \emph{PageRank}, which seem not sensitive to software stack from the perspective of load buffer full stall.
The notable one is Sort, which has about 72.5 times more load buffer full stalls when the full software stack involved.

Long latency memory accesses increase the load buffer's pressure since lots of load operations are in-flight and the new load operations can not be issued for lacking of load buffer entries, where a pipeline load buffer full stall occurs.
This phenomenon implies that the participation of full software stack increases memory access latency for most of big data analytics applications~\footnote{We also observe a similar phenomenon for store buffer. The applications run at pseudo-distributed mode have more store buffer full stalls than their standalone counterpart. However, the ratio is not as big as load buffer full stalls. The maximum ratio is about 1.6. For the store buffer full stall does not have a strong correlation with performance, we do not discuss it here.}.

\begin{figure}
\centering
\includegraphics[scale=0.9]{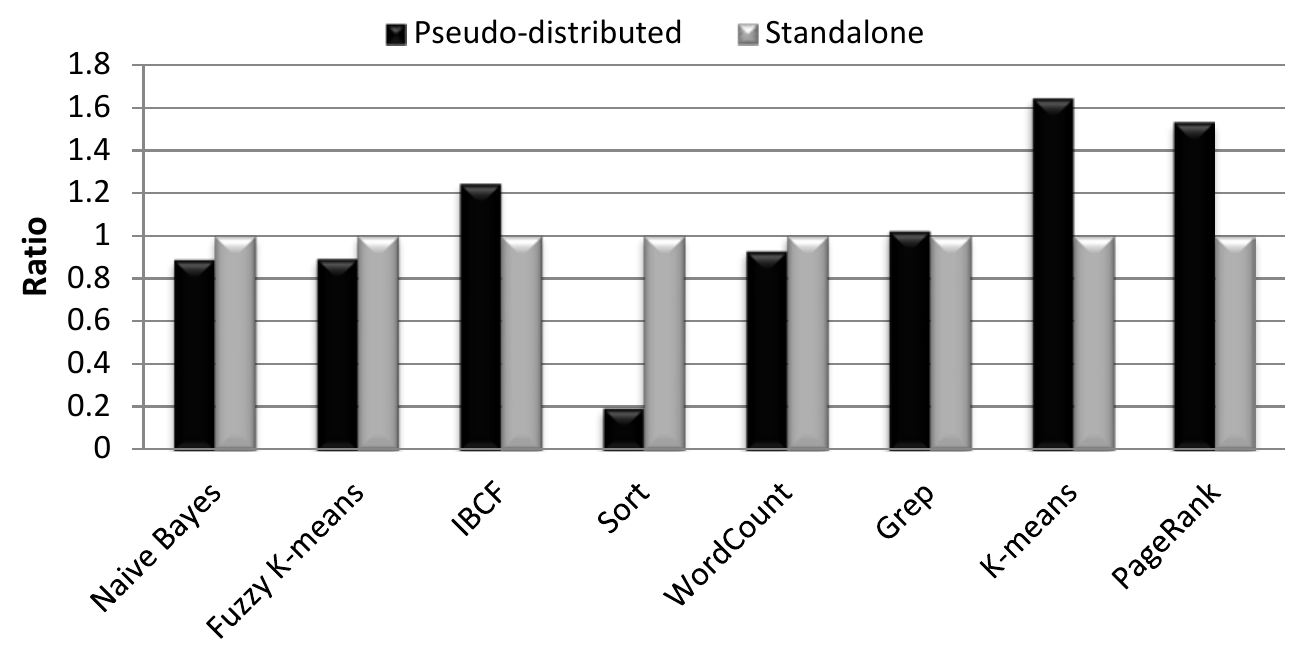}
\caption{L2 cache MPKI.}\label{local-l2}
\end{figure}

\begin{figure}
\centering
\includegraphics[scale=0.9]{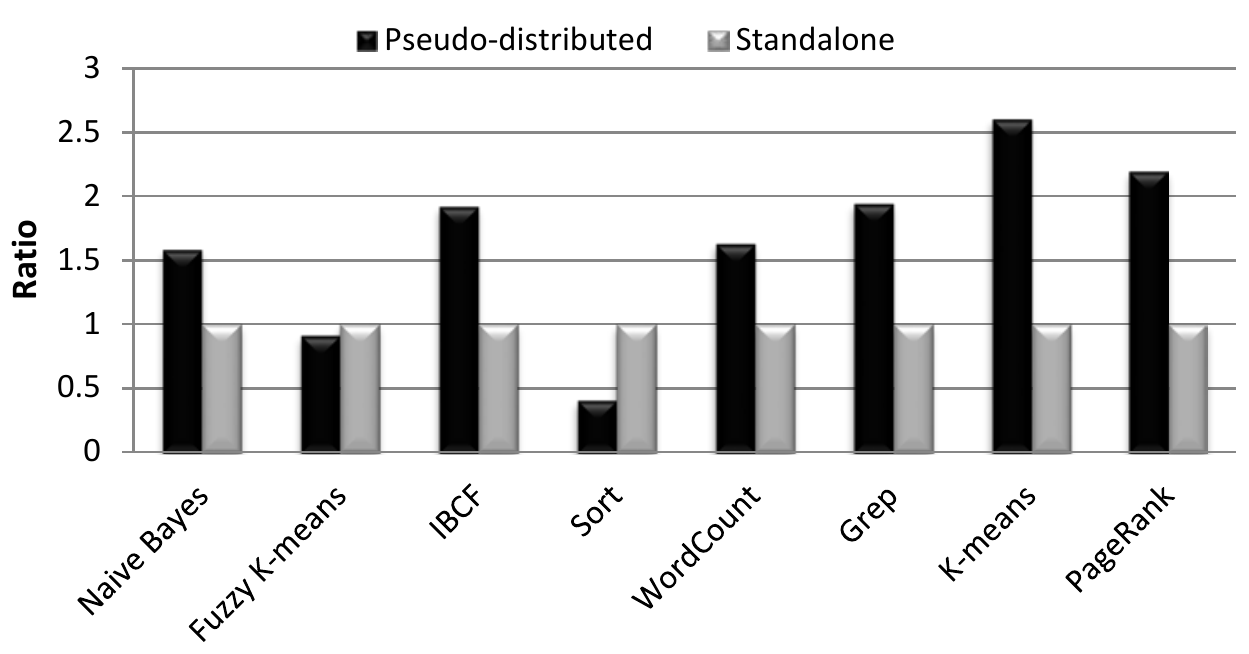}
\caption{Data TLB MPKI.}\label{local-tlb}
\end{figure}

In Section~\ref{corr} we find that the private united cache and TLB miss ratio have strong positive correlations with CPI. Also the private united cache (L2 cache) and TLB are key units along the data path.
Figure~\ref{local-l2} and Figure~\ref{local-tlb} show the L2 cache MPKI (misses per thousand instructions) and data TLB MPKI. Here we also show the normalized value using standalone mode as baseline just as previous subsection does.

We can find from Figure~\ref{local-l2} that for different applications the software stack has different impacts on L2 cache and data TLB behaviors.
For some applications, the pseudo-distributed mode has more L2 cache misses.
Such as \emph{K-means} and \emph{Pagerank} have 1.64 and 1.53 times as many L2 cache misses running on pseudo-distributed mode as running on standalone mode respectively.
Other applications have less cache misses after invoking the full software stack.
For example, psesudo-distributed mode \emph{Sort} only has 20\% L2 cache misses of its standalone counterpart.
For data TLB performance, we can find from Figure~\ref{local-tlb} that nearly all of the big data analytics applications have more data TLB misses per thousand instructions when they run under the pseudo-distributed mode.
The notable one is \emph{K-means}, which has 1.6 times more data TLB misses running at pseudo-distributed mode than running at standalone mode.
The only two exception are \emph{Sort}, which has less data TLB misses when using full software stack (only 0.41 time of that of its standalone counterpart), and \emph{Fuzzy K-means}, which seams insensitive to software stack from the perspective of data TLB behavior.
So for most of big data analytics applications, the participation of full software stack increases the burden of the data TLB.

The L2 cache miss and data TLB miss can trigger long latency memory access. And we can find that some applications have less L2 cache and less data TLB MPKI whereas own more load buffer full stalls, such as \emph{Sort}.  Those phenomena are not inconsistent.
The event of L2 cache miss or TLB miss records how many times the miss happened.
The load buffer full event records the total cycles stalled due to lack of load buffer entry.
We calculate L2 cache miss ratio and TLB miss ratio by normalizing those misses with the total number of instructions retired.
The participation of full software stack increases the total number of instructions executed by processor, which also enlarge the denominator of L2 cache MPKI and data TLB MPKI.
So the software stack instructions amortize the L2 cache MPKI and data TLB MPKI.
However the software stack does not reduce the memory access latency. On the contrary, the full software stack incurs larger working set~\footnote{We also examine the L3 cache statistic. We find that all of the big data analytics applications have more L3 cache MPKI when they are running at pseudo-distribute mode. The ratio ranges from 1.4 to 2.7. This phenomenon also indicates that the application running at pseudo-distribute mode owns larger working set. We do not show the data both for the space limitation and the L3 cache miss does not have a strong correlation with CPI performance.} and prolong the memory access latency for most of the big data analytics applications.

\subsection{Kernel-mode Instruction Ratio}
\begin{figure}
\centering
\includegraphics[scale=0.85]{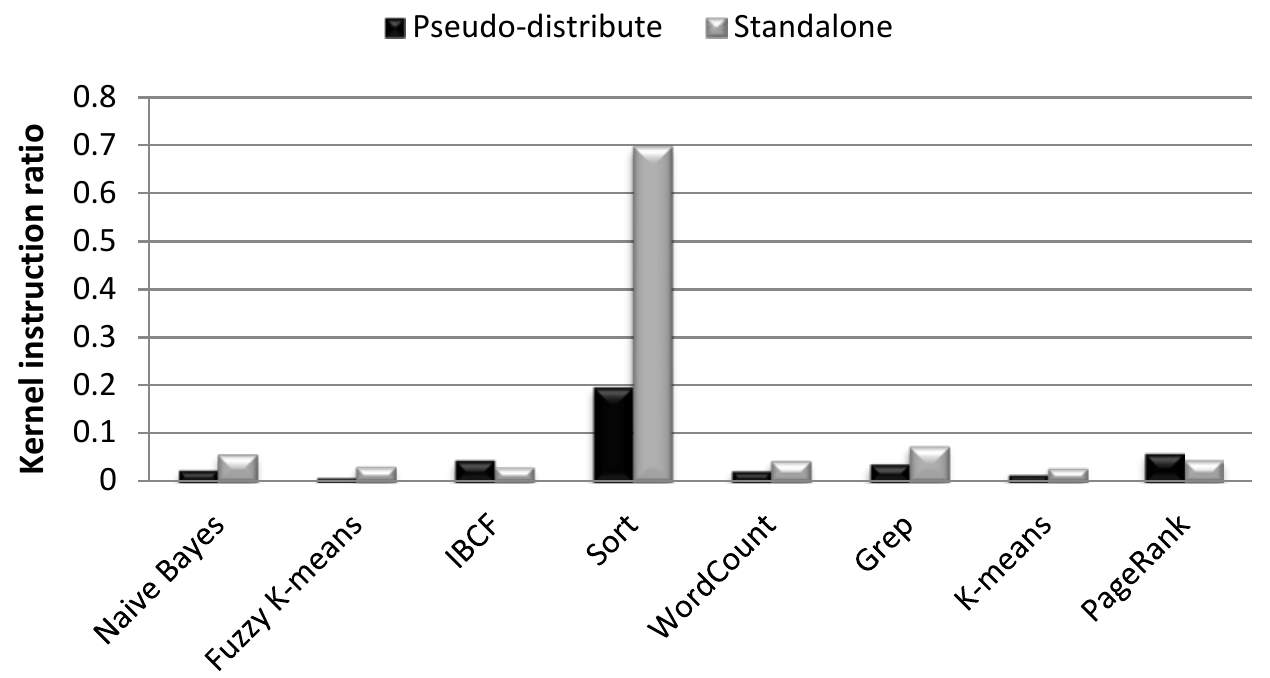}
\caption{Kernel-mode instruction ratio.}\label{local-ins}
\end{figure}

Figure~\ref{local-ins} illustrates the retired kernel-mode instruction ratio.
We can find that applications running at pseudo-distributed mode have less kernel-mode instructions than their standalone counterpart for most of data analytics workloads.
The exceptions are \emph{PageRank} and \emph{IBCF}, which have slightly increased kernel-mode instructions, no more than 2\%.
This implies that the software stack do not invoke a lot of system calls. Most of the functions are implements in the application level.
So most of the instructions introduced by full software stack are executed at user-mode (i.e. executed on ring 1 to ring 3). The kernel-mode instruction's ratio is diluted.
The notable one is \emph{Sort}, which triggers a lot of system calls as explained in Section~\ref{sub_ipc}. After invoke the full software stack, a large amount user-mode instructions  dilute the kernel-mode instruction ratio and make its kernel-mode instruction ratio be reduced from 0.7 to 0.2. 


\subsection{Observations and Implications}

From above comparative experiments, we find that the software stack has the following impacts on application behaviors from the perspective of micro-architecture.
1) The software stack increases application binary size and 
aggravates the front end inefficiency for big data analytics workloads.
2) Even though software stack can amortize cache miss ratio and TLB miss ratio for some applications, it does not decrease the memory access latency. On the contrary, the full software stack incurs larger working set and increases the memory access latency especially for load operations.
3) Most of the software stack's functions are implemented on application level and do not invoke lots of system calls. So most of the instructions are executed at user-mode and reduce the whole applications's kernel-mode instruction ratio.


In order to optimize big data analytics applications developed with the typical big data software stack, i.e. Hadoop in this paper, the potential burden introduce by third-party libraries and software stacks should be noticed, such as the data operations that may incur long memory access latency.
And the OS functions' performance should not be pay much attention for the software stack does not invoke lots of kernel-mode instructions.


\section{Conclusion}\label{conclusion}

In this paper, after investigating  most important application domains in terms of page views and daily visitors,
we chosen eleven representative big data analytics workloads and  characterized their  micro-architectural characteristics on the systems equipped with modern superscalar out-of-order processors by using hardware performance counters.



Our study on the workload characterizations reveals
that the big data analytics applications
share many inherent characteristics,
which place them in a different class from desktop, HPC, chip multiprocessors, traditional service and scale-out service workloads.
Meanwhile, we also observe  that the scale-out service workloads (four among six benchmarks in CloudSuite) share many similarities in terms of micro-architectural characteristics  with that of the traditional server workloads characterized by SPECweb 2005 and TPC-W. 

Our correlation analysis shows that
even though big data analytics workloads suffer from notable front end stalls, the factor that affects CPI performance most is not the front end stall, but the long latency data accesses. So the
long latency data accesses should be reduced with the highest priority for big data analytics applications.

Our investigation finds that the typical big data analytics software stack, i.e. Hadoop, does have impacts on application behaviors, especial on instruction fetch
unit and load operations.
So for programmers who writer big data analytics applications with the big data software stack, should pay much attention to the burden brought by the software stack.

\parskip=0pt
\parsep=0pt
\bibliographystyle{ieeetrsrt}
\bibliography{IEEEabrv,tex}


\end{document}